%% file: PFSPH_Droplets_arxiv.tex
\documentclass[12pt]{article}

\usepackage[a4paper, margin=1in]{geometry}

\newcommand{\firstRevision}[1]{{#1}}

\usepackage{newtxtext}
\usepackage{newtxmath}
\usepackage{anyfontsize}

\usepackage{natbib}
\usepackage{hyperref}
\hypersetup{
    colorlinks = true,
    urlcolor   = blue,
    citecolor  = blue,
    linkcolor = blue
}

\newcommand{\RomanNumeralCaps}[1]
\linenumbers

\input{src/0packagesAndMacros.tex}

\title{Computing sessile droplet shapes on arbitrary surfaces with a new pairwise force smoothed particle hydrodynamics model}
\author{Riley M Whebell\footremember{QUT}{School of Mathematical Sciences, Queensland University of Technology, Brisbane QLD 4000, Australia} \and Timothy J Moroney\footrecall{QUT} \and Ian W Turner\footrecall{QUT} \and Ravindra Pethiyagoda\footnote{School of Information and Physical Sciences, University of Newcastle, Callaghan NSW 2308, Australia} \and Scott W McCue\footrecall{QUT}}
\date{}

\begin{document}

\maketitle

\begin{abstract}
    \input{src/00abstract}
\end{abstract}

\section{Introduction}
\label{sec:dropletIntroduction}
\input{src/1introduction}

\section{Numerical formulation}
\label{sec:dropletMethods}
\input{src/2methods}

\section{Validation of numerical scheme}
\label{sec:dropletValidation}
\input{src/3validation}

\section{Case studies \& numerical experiments}
\label{sec:dropletResults}
\input{src/4results}

\section{Conclusion}
\label{sec:dropletConclusion}
\input{src/5conclusion}

\appendix

\section*{Supplementary data}
Code for the SPH implementation in this work is available at \url{https://github.com/rwhebell/Whebell2024_SessileDroplets}.

\section*{Acknowledgements}
We thank the associated industry partners Syngenta, NuFarm, and Plant Protection Chemistry NZ for their involvement, and collaborators Justin Cooper-White and Phil Taylor for many fruitful discussions.
Computational resources used in this work were provided by the eResearch Office, Queensland University of Technology, Brisbane, Australia.

\section*{Funding}
This work was financially supported by an Australian Research Council Research Training Program scholarship, as well as the Australian Research Council Linkage Grant LP160100707 and associated industry partners Syngenta and Nufarm.

\section*{Declaration of interests}
The authors report no conflict of interest.

\setlength{\bibsep}{0em}
\bibliography{dropletPaperRefs.bib}

\end{document}

%% file: src/0packagesAndMacros.tex
\usepackage{tikz}

\usepackage{caption}
\usepackage{subcaption}

\usepackage[section]{placeins} 

\usepackage{booktabs}

\newcommand{\footremember}[2]{%
    \footnote{#2}
    \newcounter{#1}
    \setcounter{#1}{\value{footnote}}%
}
\newcommand{\footrecall}[1]{%
    \footnotemark[\value{#1}]%
}

\usepackage{bm}

\renewcommand{\vec}[1]{\bm{\mathrm{#1}}}
\newcommand{\matderiv}[2]{\frac{\mathrm{D} #1}{\mathrm{D} #2}}
\newcommand{\eucnorm}[1]{\| #1 \|}
\newcommand{\grad}[1]{\vec{\nabla} #1}
\newcommand{\laplacian}[1]{\nabla^2 #1}
\newcommand{\dv}[2]{\frac{\mathrm{d}#1}{\mathrm{d}#2}}
\newcommand{\vdot}{\cdot}

\newcommand{\bigO}{\mathcal{O}}
\newcommand{\timestep}[1]{{\left(#1\right)}}
\newcommand{\half}{\frac{1}{2}}
\newcommand{\gradWij}{\grad_i W_{ij}}

%% file: src/00abstract.tex
The study of the shape of droplets on surfaces is an important problem in the physics of fluids and has applications in multiple industries, from agrichemical spraying to microfluidic devices. 
Motivated by these real-world applications, computational predictions for droplet shapes on complex substrates -- rough and chemically heterogeneous surfaces -- are desired. 
Grid-based discretisations in axisymmetric coordinates form the basis of well-established numerical solution methods in this area, but when the problem is not axisymmetric, the shape of the contact line and the distribution of the contact angle around it are unknown. 
Recently, particle methods, such as pairwise force smoothed particle hydrodynamics (PF-SPH), have been used to conveniently forego explicit enforcement of the contact angle. 
The pairwise force model, however, is far from mature, and there is no consensus in the literature on the choice of pairwise force profile. 
We propose a new pair of polynomial force profiles with a simple motivation and validate the PF-SPH model in both static and dynamic tests.
We demonstrate its capabilities by computing droplet shapes on a physically structured surface, a surface with a hydrophilic stripe, and a virtual wheat leaf with both micro-scale roughness and variable wettability. 
We anticipate that this model can be extended to dynamic scenarios, such as droplet spreading or impaction, in the future.

%% file: src/1introduction.tex
The task of calculating the equilibrium shape of a sessile droplet on an arbitrary surface, including the effects of gravity, is surprisingly difficult, despite the simplicity of the problem description.  
In the simplest case of a flat horizontal surface, one can proceed by solving the Young-Laplace equation for the droplet shape given the droplet volume and the prescribed contact angle \citep{danovShape2016}. 
However, even for a similar geometry such as an inclined plane, the lack of rotational symmetry leads to numerous additional complications that make the problem considerably more challenging. 
In particular, the boundary of the wetted area on the substrate (the contact line) is then completely unknown, as is the contact angle at each point on this boundary. 
Therefore, certain approximations are often made to make progress on this more difficult problem, such as assuming a circular contact line \citep{brownStatic1980,tredenickEvaporating2021}, or using empirical models for the contact angle distribution around the contact line \citep{raviannapragadaDroplet2012}. 
For a more general surface geometry, the challenges are even more stark, as deviations in surface topology lead to a highly nontrivial formulation again with an unknown contact line location and unknown contact angles. 

A related problem is to study the evolving shape of a droplet that has been released on a surface with a relatively low energy field, and determine the dynamics of the droplet shape as it settles towards equilibrium. 
In addition to the inherent challenges described above for the sessile droplet, this time-dependent problem is further complicated by the unknown relationship between the speed of the contact line and the contact angle \citep{hockingRival1992}.
For our purposes, we are interested in using this time-dependent framework as a means to compute shapes of sessile droplets on arbitrary surfaces.

Some promise is shown in this area by energy minimisation methods, although these methods do not incorporate viscous effects, or model the temporal behaviour of settling droplets \citep{jamaliNumerical2021}.
In contrast, in this study, we develop a numerical method based on smoothed particle hydrodynamics (SPH) to tackle these challenging problems. 
In the context of computing shapes of sessile droplets, one advantage of our approach is that the geometry of the droplet at the contact line arises naturally as a consequence of the SPH formulation, rather than as an input to the model.

SPH is a computational method for discretising fluid flow problems using ``particles'' that carry information about the fluid, originally developed by \cite{gingoldSmoothed1977} and \cite{lucyNumerical1977}; and more recently reviewed by \cite{wangOverview2016} and \cite{yeSmoothed2019}, for example.
The particles serve as interpolation nodes for fluid properties such as density and velocity. The particles are not stationary and not connected; rather, they are advected with the flow according to the fluid velocity field $\vec{v}(\vec{x}, t)$. 
This Lagrangian discretisation transforms partial differential equations for any given field value $\varphi(\vec{x}, t)$ into ordinary differential equations for each particle's value $\varphi_i(t)$, coupled with the advection equation $\mathrm{d}\vec{x}_i/\mathrm{d}t = \vec{v}_i$ for the particle's position $\vec{x}_i$. 
Provided that sufficiently many particles are used in the simulation, the particle properties can then be interpolated to reconstruct the full fields everywhere in the domain.

Since its inception, SPH has been applied to a wide variety of problems, including astrophysics \citep{springelSmoothed2010}, coastal engineering \citep{barreiroSmoothed2013}, porous media flow \citep{zhuSmoothed2001}, and even blood flow in injured arteries \citep{mullerInteractive2004}.
The key advantage of SPH in simulating droplet behaviour is that the scheme does not necessitate the tracking of the sharp interface between the liquid and the air. 
Indeed, it does not track the interface directly at all; instead, the interface is deduced \textit{a posteriori} based on the density field. 

\firstRevision{
Some practitioners of SPH introduce surface tension effects by means of the continuum surface force (CSF). 
This idea dates back to \cite{brackbillContinuum1992} (in general) and \cite{morrisSimulating2000} (applied to SPH). 
The CSF method starts with the Young-Laplace pressure boundary condition at the liquid-gas interface, but translates the normal force per unit area on the interface into a force per unit volume throughout the liquid, by means of a smoothing function. 
That is, a force $\vec{F}_\mathrm{CSF}$ is introduced everywhere in the fluid:
\[
    \vec{F}_\mathrm{CSF} = \delta_\mathrm{CSF}(\vec{x}) \, \sigma K(\vec{x}) \hat{\vec{n}}(\vec{x}),
\]
where $\delta$ is the smoothing function (maximised at the interface, and vanishing away from it), $\sigma$ is the surface tension, $K = -\nabla \cdot \hat{\vec{n}}$ is the curvature of the interface, and $\hat{\vec{n}}(\vec{x})$ is the unit normal of the interface, extended into the fluid \citep{morrisSimulating2000}.
The success of these methods hinges on robust calculation of the smoothing function and the unit normal `field'.
Recent approaches \citep{vergnaudCCSF2022,blankSurface2024,antuonoCoalescing2025} require kernel corrections, which increase the computational burden of the method significantly, and smoothing of the normal field, as well as special corrections for thin jets.
The matter is further complicated when one considers contact angle effects for droplets on substrates -- \cite{chironFast2019} resort to tracking the liquid-solid interface explicitly.
This additional complexity yields accurate results; however, our aim is to produce a robust method applicable to droplet simulations on arbitrarily rough substrates.
As such, we seek a method that is more computationally efficient and robust to interfaces with high curvatures, which we expect to be prevalent in our application.
}

We choose the SPH method because it provides the needed flexibility to model droplets with non-trivial shapes on complex substrates that would otherwise pose a significant challenge for interface tracking or fixed grid methods \citep{yeSmoothed2019}.
In the spirit of this interface-free approach, in the past decade a modified SPH method has emerged in which the Young-Laplace pressure boundary condition at the free surface is replaced with inter-particle forces that mimic cohesion and adhesion, in what is known as the pairwise force SPH method (PF-SPH). 

\cite{kordillaSmoothed2013}, \cite{tartakovskyPairwise2016}, and \cite{shigorinaSmoothed2017} all used a PF-SPH method to study droplet shapes on rough surfaces, although each used a different function for the pairwise force. 
\cite{santacruz-yungaPair2025} use a potential based on the popular Lennard-Jones potential for the pairwise interactions, which in this framework is analogous to a pairwise force.
There is, in general, a lack of consensus regarding the best pairwise force formulation, and a lack of understanding about what effect the formulation has on the simulated surface tension and contact angle.
Existing pairwise force profiles in the literature are rarely physically motivated and seldom investigated in their own right.
\firstRevision{
Nevertheless, the PF-SPH method is very attractive for its simplicity, computational efficiency, and robustness to thin features of both substrates and droplets. 
}
With this in mind, our main contribution is to propose a new force profile for PF-SPH that is physically motivated, scale-independent, and carefully validated through the reproduction of multiple independent physical phenomena.
To this end, we have developed a new method for calibrating the contact angle on an ideal surface by fitting whole droplet shapes obtained from semi-analytical solutions to the Young-Laplace equation.
With the new PF-SPH model thoroughly calibrated and validated, we then demonstrate its application to several important test problems from the literature.
First, we calculate the equilibrium shape of settling droplets on microscopically rough and chemically patterned substrates.
Then, we apply the model to simulate a scenario from an agricultural application: a droplet impacting and settling on a virtual plant leaf.

The paper is structured as follows. 
In Section \ref{sec:dropletMethods} we outline the details of our weakly compressible pairwise-force SPH model. 
We describe a new, scale-independent, pairwise force term, and give some details on boundary handling, time integration, and computational implementation.
Several validation tests are carried out in Section \ref{sec:dropletValidation}, in which we ensure the model reproduces expected interfacial phenomena in both static and dynamic scenarios.
In particular, in Section \ref{sec:YLprofiles} we detail our method of calibrating the pairwise force to semi-analytical solutions for droplet shapes on flat surfaces.
Section \ref{sec:dropletResults} is then devoted to some numerical experiments of interest.
Firstly, we simulate two small ($3\mu\mathrm{L}$) droplets with different initial velocities on a surface with regular, sharp, square pillars, reproducing the experimental results of \cite{dupuisModeling2005} and showing a transition between wetting states. 
Next, we simulate a droplet settling on a flat, hydrophobic, surface with a hydrophilic stripe, and observe a smooth transition in the droplet's contact angle from the equilibrium hydrophobic contact angle to the equilibrium hydrophilic contact angle.
We then simulate a droplet settling on a virtually reconstructed wheat leaf surface in the context of broader work on understanding spray-canopy interactions on plants \citep{dorrSpray2016, dorrImpaction2015, mayoSimulating2015, tredenickEvaporating2021}.
We note that our PF-SPH scheme is quite flexible and should be applicable to a broad range of time-dependent droplet-related problems on complex substrates such as droplet impaction and spreading.
Finally, the \texttt{julia} code containing our implementation of the model is \href{https://github.com/rwhebell/Whebell2024_SessileDroplets}{available online} on GitHub.

%% file: src/2methods.tex
\subsection{Governing equations}
The continuum model we use is the weakly compressible, barotropic Navier Stokes equations
\begin{align}
    \matderiv{\rho}{t} &= -\rho (\nabla \cdot \vec{v}),
    \label{eqn:continuity}
    \\
    \matderiv{\vec{v}}{t} &= \frac{-1}{\rho} \grad{P} + \frac{\mu}{\rho} \laplacian{\vec{v}} + \vec{g} + \vec{F}^\mathrm{(pf)},
    \label{eqn:momentum}
\end{align}
with $\vec{x}$ the position, $\vec{v}$ the velocity, $\rho$ the density, $P$ the pressure, $\mu$ the dynamic viscosity, and $\vec{g}$ the gravitational acceleration.
The main focus of this work will be the body force (per unit mass) $\vec{F}^\mathrm{(pf)}$, which we will construct in Section \ref{sec:pairwiseForceModel} to reproduce surface tension and contact angle effects.
The derivative $\mathrm{D}/\mathrm{D}t$ is the Lagrangian derivative
\[
\matderiv{}{t} = \frac{\partial}{\partial t} + \vec{v} \cdot \vec{\nabla},
\]
which, as we shall see, lends itself naturally to discretisation using smoothed particle hydrodynamics.

\subsection{Discretisation}
As is standard in SPH methods, we represent the fluid by $N$ particles, each with their own position, velocity, density, and label (to distinguish fluid from solid, for example). 
Using $i$ and $j$ as particle indices, the discretised SPH form of the model \eqref{eqn:continuity}-\eqref{eqn:momentum} is
\begin{align}
    \dv{\rho_i}{t} &=
    \rho_i \sum_{j=1}^N ( \vec{v}_{ij} \vdot \gradWij ) V_j,
    \label{eqn:SPHcontinuity}
    \\
    \dv{\vec{v}_i}{t} &= 
    \frac{-1}{\rho_i} \sum_{j=1}^N (P_i + P_j - \Pi{}) \gradWij V_j
    + \vec{g}
    + \vec{F}_i^\mathrm{(pf)},
    \label{eqn:SPHacceleration}
    \\
    \dv{\vec{x}_i}{t} &= \vec{v}_i,
    \notag
    \\
    \Pi &= 2(d+2)\mu \frac{\vec{v}_{ij} \vdot \vec{x}_{ij}}{\eucnorm{\vec{x}_{ij}}^2},
    \notag
    \\
    P_i(\rho_i) &= \frac{c_0^2 \rho_0}{7} \left[ \left( \frac{\rho_i}{\rho_0} \right)^7 - 1 \right],
    \label{eqn:SPH_EOS}
\end{align}
where
$\vec{x}_{ij} = \vec{x}_i - \vec{x}_j$ for notational convenience,
$W_{ij} = W(\vec{x}_{ij}; H)$ is an SPH kernel with compact support radius $H$, 
$V_j = m_j / \rho_j $ is the volume of particle $j$,
and
$d$ is the spatial dimension (always equal to 3 in this work).
The pressure $P_i$ is calculated from the density $\rho_i$ with the equation of state \eqref{eqn:SPH_EOS}.
The constants $c_0$ and $\rho_0$ are the artificial speed of sound and the reference density, respectively.
Note that we have switched from using the material derivative $\mathrm{D}/\mathrm{D}t$ to the total derivative $\mathrm{d}/\mathrm{d}t$, as the time derivative of a particle's property follows the flow by definition.

In this work we use the quintic Wendland function \citep{wendlandPiecewise1995} as the SPH kernel, which we parameterise by $H$, its radius of support, also called the kernel cutoff radius:
\begin{align*}
    W(\vec{x} - \vec{x}'; H) &= 
    \frac{21}{2 \pi H^3} \; w(\eucnorm{\vec{x} - \vec{x}'} / H),
    \\
    w(\tau) &= 
    \begin{cases}
        (1 - \tau)^4 (4\tau + 1), & 0 \leq \tau < 1, \\
        0, & \tau \geq 1.
    \end{cases}
\end{align*}
Wendland functions were developed independently of SPH for their smoothness properties, but have been found to give accurate and stable SPH results \citep{dehnenImproving2012}.
In the literature, this kernel is sometimes parameterised by its smoothing length $h$, which is defined as half the kernel cutoff radius.
To avoid this confusion, we parameterise the kernel by its cutoff, which we denote $H$, and set $H / \Delta x = \kappa = 4$, where $\Delta x$ is the particle width.

With the exception of the pairwise force term $\vec{F}^\mathrm{(pf)}$, which will be the main focus of this work, the discretisation summarised above is well established in the literature: see, for example, \cite{monaghanSmoothed2005}.
In this work we will implement and validate a new form of the pairwise force term to model surface tension and contact angle effects, and highlight some particular properties of $\vec{F}^\mathrm{(pf)}$ that yield stable and accurate simulations of droplets.

\subsubsection{Continuity equation}
The operator we use for the divergence of velocity is from \cite{monaghanSmoothed2005}:
\begin{align*}
    \nabla \cdot \vec{v} &=
    - \sum_{j=1}^N ( \vec{v}_{ij} \vdot \gradWij ) V_j,
\end{align*}
and is specifically constructed to vanish when $\vec{v}$ is constant.

\subsubsection{Pressure gradient \& equation of state}
The discretisation of the gradient of pressure that we use, namely
\begin{align*}
    (\vec{\nabla} P)_i
    &=
    \sum_{j=1}^N (P_i + P_j) \vec{\nabla}_i W_{ij} V_j,
\end{align*}
is due to \cite{bonetVariational1999} (and recently used by \cite{dominguezDualSPHysics2022}), who showed it to be consistent with equation \eqref{eqn:SPHcontinuity} and to conserve linear momentum exactly.
Note that since $\vec{\nabla}_i W_{ij} = -\vec{\nabla}_j W_{ji}$, we have the anti-symmetry $(\vec{\nabla} P)_i = -(\vec{\nabla} P)_j$.
This is the property that conserves momentum, since the contribution of particle $j$ to ${\mathrm{d}\vec{v}_i}/{\mathrm{d}t}$ is equal and opposite to the contribution of particle $i$ to ${\mathrm{d}\vec{v}_j}/{\mathrm{d}t}$.

The pressure is calculated from the density by the equation of state. 
We follow \cite{monaghanSimulating1994,monaghanSmoothed2005} in using
\begin{align*}
    P_i &= \frac{c_0^2 \rho_0}{7} \left[ \left( \frac{\rho_i}{\rho_0} \right)^7 - 1 \right],
\end{align*}
which was originally reported by \citet[p. 44]{coleUnderwater1948} in the study of underwater explosions. 
\citeauthor{coleUnderwater1948} notes that they chose the exponent $7$ to approximately fit experimental data. 
In our case, we find that the results are not at all sensitive to the particular equation of state in use -- even the first-order approximation $P(\rho) = c_0^2 (\rho - \rho_0)$ gives almost indistinguishable results.
The coefficient $c_0$ is an artificial speed of sound that controls the compressibility of the fluid. 
The actual speed of sound in water is around $1500\mathrm{m/s}$, but such a value would require the use of extremely small timesteps to properly resolve pressure waves travelling at that speed.
Instead, in the weakly compressible model, we use an artificial value of $c_0$ on the order of $100\;\mathrm{m/s}$ such that density fluctuations are kept to within $1\%$ of the reference value $\rho_0$ \citep{monaghanSmoothed2005}.
The artificial speed of sound required for a particular problem can be estimated as $10 v_\text{max}$, an order of magnitude greater than the maximum expected fluid speed.

\subsubsection{Viscosity}
The discrete operator we use to approximate the Laplacian is that of \cite{monaghanShock1983}, namely
\begin{align*}
    (\laplacian{\vec{v}})_i &= 
    2(d+2) \sum_{j=1}^N \frac{\vec{v}_{ij} \vdot \vec{x}_{ij}}{\eucnorm{\vec{x}_{ij}}^2} \gradWij V_j,
\end{align*}
where $d$ is the spatial dimension.
More recently proposed operators exist with favourable properties, but we have chosen this particular discretisation based on the analysis by \cite{colagrossiTheoretical2009} that suggests it is more appropriate for the simulation of free surfaces.

\subsubsection{Solid boundary treatment}
At the fluid-solid interface, we have the no-slip condition: $\vec{v} = \vec{0}$.
We impose this condition indirectly, representing the solid substrate with fixed dummy particles, initialised on a regular grid with spacing $\Delta x$.
Multiple layers of these particles are used to avoid an SPH neighbourhood deficiency at the boundary.
The solid particles have the same physical properties as the fluid at rest, with mass $\rho_0 \Delta x^3$ and density $\rho_0$, but with zero velocity.
\firstRevision{
This simple approximation of the no-slip condition has been shown to be consistent in the limit as $H\rightarrow 0$, although does not reproduce it exactly in simulations.
Other schemes utilise mirroring approaches \citep{maciaTheoretical2011}, in which fluid properties are interpolated and extended into the solid boundary using surface normals. 
In our study of droplets on rough surfaces, however, this approach would be quite complex to implement for a general surface. 
For example, it is not straightforward to map any point in space to its closest surface point and thus the appropriate surface normal.
Furthermore, some boundary points may be assigned incorrect properties because their associated interpolation point above the solid is in the gas phase, which we do not discretise with SPH particles.
The subject of more accurate boundary condition treatment is thus left to future work, and the interested reader is referred to \cite{maciaTheoretical2011} for more details.
}

\begin{figure}
    \centering
    \includegraphics[width=\textwidth]{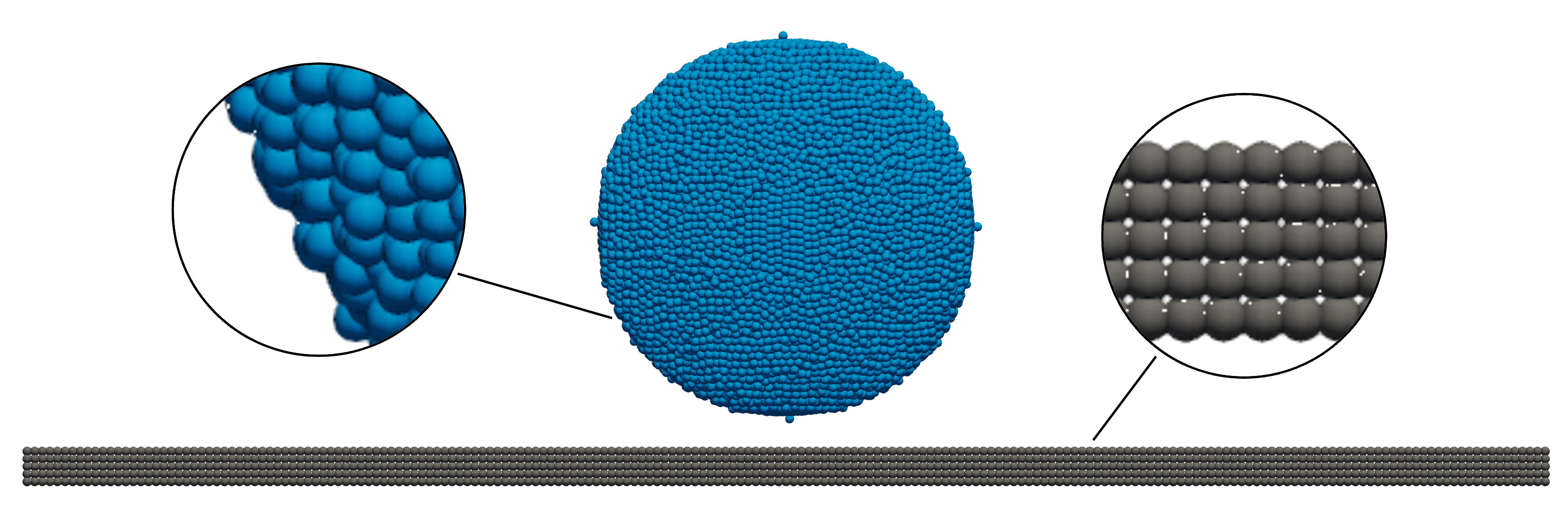}
    \caption{A side view of a 3D particle simulation in which a droplet of fluid particles (blue) is about to impact solid boundary particles (grey). Insets show closeups of each type of particle. Fluid particles are advected with the flow and thus disorganised, while solid particles are fixed on a cubic lattice. The particles are rendered as spheres of volume $m_j / \rho_j$, but we note that this is for visualisation only: in the SPH scheme, they are better understood as interpolation nodes.}
    \label{fig:CAcal_setup}
\end{figure}

The ``dummy'' solid particles are included in the summations over $j$ in the discrete continuity and momentum equations \eqref{eqn:SPHcontinuity} and \eqref{eqn:SPHacceleration}, as if they were fluid particles.
Pressure forces and the repulsion due to pairwise forces ensure that fluid particles do not penetrate the boundary.
Figure \ref{fig:CAcal_setup} shows a typical setup in which a droplet of fluid particles is about to impact a flat solid boundary. 

\subsubsection{Time integration}
The time integration scheme we use is a second-order, symplectic, position-Verlet scheme. 
It is modified from \cite{leimkuhlerMolecular2015} by \cite{dominguezDualSPHysics2022} to include a velocity half-step, which is necessary to integrate the continuity equation and calculate the viscous term.
We will use the following notation for the discretised governing equations \eqref{eqn:continuity} and \eqref{eqn:momentum}:
\[
    \dv{\rho_i}{t} := Q_i, \quad \dv{\vec{v}_i}{t} := \vec{A}_i.
\]
Dropping particle indices for clarity, and letting the value of a quantity $\varphi$ at timestep $k$ be denoted $\varphi^{(k)}$, the original position Verlet scheme \cite[p. 107]{leimkuhlerMolecular2015} reads
\begin{align*}
    \vec{x}^\timestep{k+\half} &= 
    \vec{x}^\timestep{k} + \frac{\Delta t}{2} \vec{v}^\timestep{k},
    \\
    \vec{v}^\timestep{k+1} &=
    \vec{v}^\timestep{k} + \Delta t \, \vec{A}^\timestep{k + \half},
    \\
    \vec{x}^\timestep{k+1} &= 
    \vec{x}^\timestep{k+\half} + \frac{\Delta t}{2} \vec{v}^\timestep{k+1}.
\end{align*}
However, the viscous term in $\vec{A}^\timestep{k + \half}$ involves the velocity at the half timestep, so \cite{dominguezDualSPHysics2022} introduce the intermediate step
\[
    \vec{v}^\timestep{k+\half} = 
    \vec{v}^\timestep{k} + \frac{\Delta t}{2} \vec{A}^\timestep{k}.
\]
Simplifying the expression for $\vec{x}^\timestep{k+1}$, and including the integration of the continuity equation, the full scheme reads
\begin{gather*}
    \text{Calculate } \vec{A}^\timestep{k} \text{ and } Q^\timestep{k},
    \\
    \vec{x}^\timestep{k+\half} = 
    \vec{x}^\timestep{k} + \frac{\Delta t}{2} \vec{v}^\timestep{k},
    \\
    \vec{v}^\timestep{k+\half} = 
    \vec{v}^\timestep{k} + \frac{\Delta t}{2} \vec{A}^\timestep{k},
    \\
    \rho^\timestep{k+\half} = 
    \rho^\timestep{k} + \frac{\Delta t}{2} Q^\timestep{k},
    \\[1em]
    \text{Calculate } \vec{A}^\timestep{k+\half} \text{ and } Q^\timestep{k+\half},
    \\
    \vec{v}^\timestep{k+1} =
    \vec{v}^\timestep{k} + \Delta t \vec{A}^\timestep{k + \half},
    \\
    \vec{x}^\timestep{k+1} = 
    \vec{x}^\timestep{k} + \frac{\Delta t}{2} \left[ \vec{v}^\timestep{k} + \vec{v}^\timestep{k+1} \right],
    \\
    \rho^\timestep{k+1} = \rho^\timestep{k} + \Delta t Q^\timestep{k+\half}.
\end{gather*}

The timestep is chosen adaptively according to viscous, maximum force, and acoustic constraints:
\begin{gather}
    \Delta t_v = \frac{\Delta x^2 \rho_0}{\mu},
    \quad
    \Delta t_a = 0.15 \sqrt{ \frac{\Delta x}{\max_j \eucnorm{\vec{A}_j}} },
    \quad
    \Delta t_c = 0.15 \frac{\Delta x}{c_0},
    \label{eq:timesteps}
    \\
    \Delta t = \min(\Delta t_v, \Delta t_a, \Delta t_c). \notag
\end{gather}
In the present application, the acoustic constraint determining $\Delta t_c$ is almost always far smaller than the other two in \eqref{eq:timesteps}, due to the small scale of the droplets and the comparatively large artificial speed of sound $c_0$.

\subsection{Implementation details}
The Lagrangian nature of SPH, while making it a very flexible method, also makes it challenging to implement efficiently. 
We follow the work of \cite{ihmsenParallel2011} in the use of some key data structures and parallel methods, which we will briefly summarise here.

Implemented naively, each sum in the discretised equations \eqref{eqn:SPHcontinuity} and \eqref{eqn:SPHacceleration} has a time complexity of $\bigO(N^2)$.
This is made more efficient by pre-computing a neighbour list
\[
\mathcal{J}(i) = \left\{ j : \eucnorm{\vec{x}_i - \vec{x}_j} < H \right\}
\]
for each fluid particle $i$.
Any sum over $j = 1,\dots,N$ then becomes a sum over $j \in \mathcal{J}(i)$.
Since density fluctuations are small in this weakly compressible scheme, the number of neighbours is approximately constant (around $4\pi\kappa^3/3$).
Thus, the complexity of calculating the particle interactions becomes $\bigO(N)$.

We accelerate the neighbour search by using a background grid with cells of width $H$. 
Each grid cell is uniquely identified by a tuple of integer coordinates $(c_1, c_2, c_3)$, and we keep a hash table of lists of particles contained in each cell.
To minimise memory allocations, we pre-allocate enough storage for each of these lists to contain $\kappa^d$ indices.
Density fluctuations are low, so the maximum number of fluid particles that one cell may contain is roughly constant.
The neighbour search then only considers particles in neighbouring cells (the number of which is roughly constant), therefore reducing the time complexity to $\bigO(N)$.

Finally, we make use of shared memory parallelism with many CPU cores wherever possible. 
The particle-particle interactions are calculated in parallel, as are the neighbour lists.
For more optimisation details we refer the interested reader to \cite{ihmsenParallel2011} for CPU implementations, or for GPU implementations: \cite{dominguezNeighbour2011,crespoGPUs2011,crespoDualSPHysics2015,dominguezDualSPHysics2022}.
Simulations were run using up to 64 processors in parallel.

\subsection{Pairwise force model for interfacial phenomena}
\label{sec:pairwiseForceModel}
Surface tension and wetting are both effects of intermolecular forces \citep{bormashenkoPhysics2017}.
A molecule at an interface misses approximately half of its interactions with neighbours when compared to a molecule in the bulk, and the resulting imbalance of forces leads to free surface energy. 
For surface tension, the relevant forces are cohesive (fluid-fluid interactions) while, for wetting, the relevant forces are adhesive (fluid-solid interactions).
Depending on the relative strengths of these forces, a droplet could be almost spherical, spread to completely wet a solid, or any configuration in between, to minimise the free surface energy. 

We aim to reproduce this behaviour on a much larger scale by mimicking intermolecular forces between SPH particles.
This is conceptually similar to the Lennard-Jones potentials used in molecular dynamics simulations \citep{jonesDetermination1924}.
The basis for these forces in SPH is empirical; nevertheless, in Section \ref{sec:dropletValidation} we will show that the pairwise force SPH model reproduces interfacial phenomena consistently and predictably.

The pairwise particle interaction forces are included in the momentum equation \eqref{eqn:SPHacceleration} via the term $\vec{F}_i^\mathrm{(pf)}$: 
\begin{equation}
    \vec{F}_i^\mathrm{(pf)} = \frac{H}{m_i} \sum_{j=1}^N s_{ij} f_{ij}( \eucnorm{\vec{x}_{ij}} / H ) \frac{\vec{x}_{ij}}{\eucnorm{\vec{x}_{ij}}},
    \label{eqn:SPHpairwiseForce}
\end{equation}
where $s_{ij}$ controls the strength of the pairwise force between particles $i$ and $j$.
Following the analogy with intermolecular potentials, we construct the pairwise force profile $f_{ij}( \eucnorm{\vec{x}_{ij}} / H )$ to be repulsive at short distances (less than one particle width), attractive at medium distances, and vanish outside the SPH kernel support radius, $H$.
In other works, this term has been constructed as a combination of Gaussians \citep{tartakovskyPairwise2016}, SPH kernels \citep{shigorinaSmoothed2017,kordillaSmoothed2013}, or part of a cosine curve \citep{tartakovskyModeling2005,nairDynamic2018}.
For simplicity, we will instead use polynomials for the force profile, with some intuitive constraints at key points.
\firstRevision{
Those constraints are:
\begin{align*}
    &f_{ij}(0) = 1, \tag{avoid trivial solution} \\
    &f_{ij}'(0) = f_{ij}'(1) = 0, \tag{smoothness at endpoints} \\
    &f_{ij}(1) = 0, \tag{continuity at kernel support radius} \\
    &f_{ij}(\kappa^* \Delta x / H) = 0. \tag{repulsion-to-attraction point}
\end{align*}
The last constraint, controlling the zero crossing of the force profile, is intentionally different for the fluid-fluid ($\mathrm{ff}$) and the fluid-solid ($\mathrm{fs}$) interactions. 
We choose $\kappa^*$ to be $1.1$ and $2$ for the fluid-fluid and the fluid-solid interactions, respectively. 
The fluid-fluid force profile is attractive at a shorter distance than the fluid-solid to prioritise cohesion over adhesion for stability at the contact line.}

\firstRevision{
We have chosen the zero crossing of the fluid-fluid pairwise force curve to be approximately the `rest distance' of the particles. 
To do this, we imagine optimally packed spheres in a face-centred cubic layout.
The Voronoi diagram of this arrangement is the rhombic dodecahedral honeycomb. 
If we equate the volume of one of our particles ($(\Delta x)^3$) with the volume of a rhombic dodecahedron, we find that the distance between neighbouring particles in such a packing is approximately $1.1 \Delta x$.
This is the geometric motivation behind the location of the zero crossing of the force profile in \eqref{eqn:SPHpairwiseForce}. 
}

\begin{figure}
    \centering
    \includegraphics[width=0.7\textwidth]{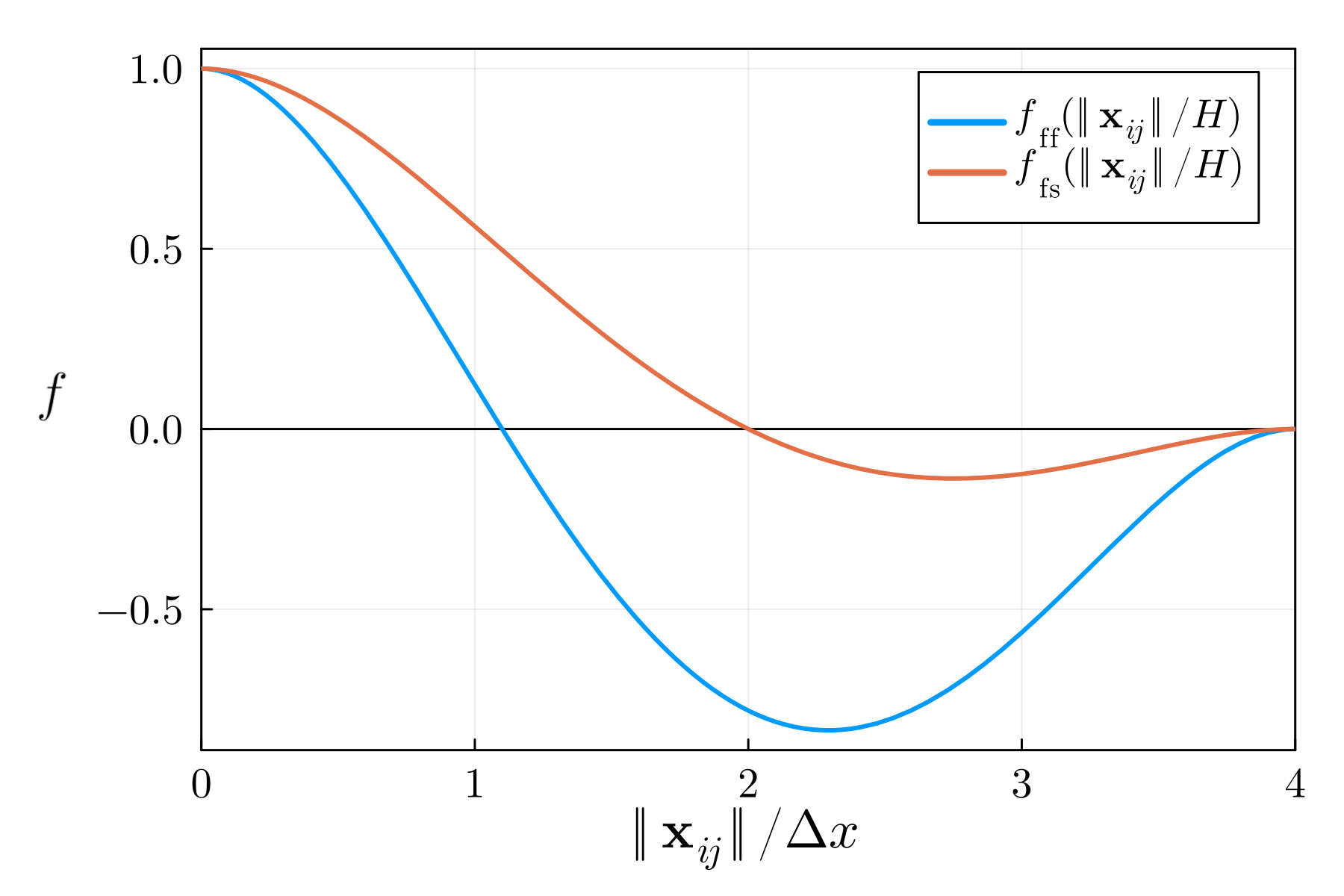}
    \caption{Distance-dependent pairwise force profiles $f_\mathrm{ff}$ and $f_\mathrm{fs}$ over the kernel support, $\tau \in [0,1]$ or $\|\vec{x}_{ij}\| / \Delta x \in [0,\kappa]$. Positive values indicate repulsion and negative values indicate attraction. Note that we have plotted the force profiles with respect to the particle width $\Delta x$ rather than the dimensionless $\tau$, to highlight the physically motivated choice of zero crossing discussed in Section \ref{sec:pf_profiles}.}
    \label{fig:pfPlot}
\end{figure}

\firstRevision{
The minimum degree of a polynomial satisfying our five constraints is four, and so we have:
\begin{gather}
    f_{ij}(\tau) = 
    \begin{cases}
        1 + \alpha_{ij}\tau^2 - 2(2 + \alpha_{ij})\tau^3 + (3 + \alpha_{ij})\tau^4, & 0 \leq \tau \leq 1,
        \\
        0, & \tau > 1,
    \end{cases}
\end{gather}
where we use $\alpha_\mathrm{ff} = -23.5$ for fluid-fluid interactions, and $\alpha_\mathrm{fs} = -11$ for fluid-solid interactions, to satisfy the zero crossing constraints above.
Figure \ref{fig:pfPlot} shows these force profiles over the range of possible particle distances.
}

The parameter $s_{ij}$ in equation \eqref{eqn:SPHpairwiseForce} controls the strength of the pairwise force and, like $f_{ij}(\tau)$, depends on the labels of the particles $i$ and $j$: fluid or solid.
We denote the cohesive force strength $s_\mathrm{ff}$ (fluid-fluid) and the adhesive force strength $s_\mathrm{fs}$ (fluid-solid).
Note the inclusion of the length scale $H$ in the form of $\vec{F}_i^\mathrm{(pf)}$ in equation \eqref{eqn:SPHpairwiseForce}: this is a departure from established pairwise-force methods \citep{kordillaSmoothed2013,nairDynamic2018,shigorinaSmoothed2017,tartakovskyPairwise2016,tartakovskyModeling2005,araiComparison2020,liuDissipative2006}, which have a scale dependency on $H$. 
Our approach ensures that the units of $s$ are that of an interfacial tension.
Taking $f_{ij}(\tau)$ to be dimensionless, the units of equation \eqref{eqn:SPHpairwiseForce} are
\begin{align*}
    [\mathrm{ms}^{-2}] &= [\mathrm{m}] [\mathrm{kg}^{-1}] s_{ij} [1] [\mathrm{m}] [\mathrm{m}^{-1}], \\
    s_{ij} &= [\mathrm{Nm}^{-1}].
\end{align*}
Thus, the units of $s$ are $\mathrm{Nm^{-1}}$, analogous to the true surface tension $\sigma$, which ensures the simulated surface tension is independent of the resolution.

\firstRevision{
While the PF-SPH model may seem an oversimplification of the complex (and small-scale) behaviour that is molecular cohesion and adhesion, it is best thought of as a coarse approximation of these phenomena, at the resolution of the smoothed particle hydrodynamics discretisation.
We will show in the next section that the resulting surface tension is linearly related to the pairwise force strength $s_\mathrm{ff}$, and that this simple model captures complex behaviour in a predictable way.
}

%% file: src/3validation.tex
In this section we present three validation tests to verify that the pairwise force model correctly reproduces surface tension and contact angle effects.
These tests not only validate the model's ability to reproduce interfacial phenomena, they also serve to calibrate the model parameters $s_\mathrm{ff}$ and $s_\mathrm{fs}$ to the material properties of interest -- the surface tension $\sigma$ (a property of the liquid) and equilibrium contact angle $\theta_\mathrm{CA}$ (a property of the liquid-substrate pair).

\subsection{Laplace pressure}\label{sec:laplace}
\begin{figure}
    \centering
    \includegraphics[width=0.7\textwidth]{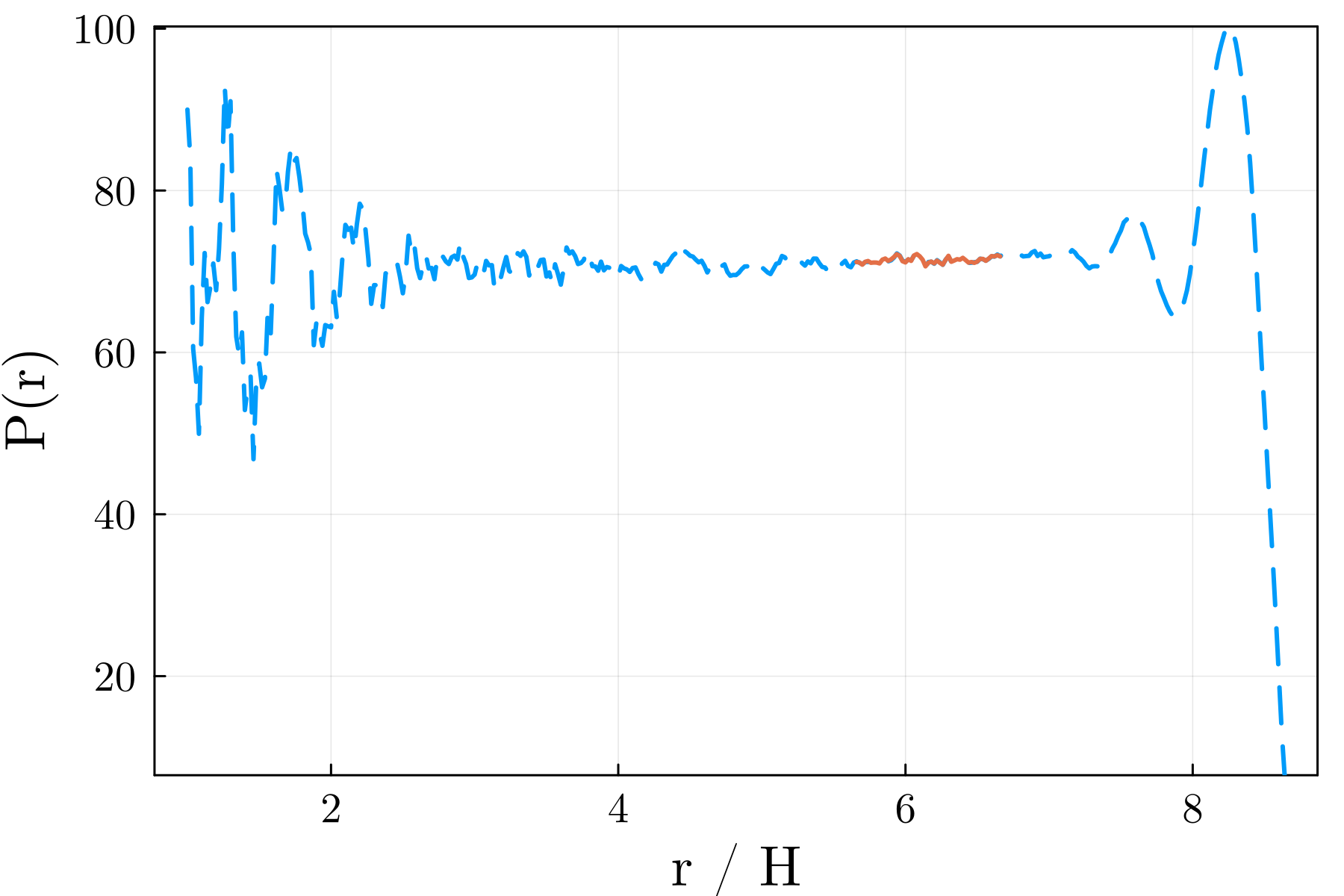}
    \caption{Virial pressure calculated using equation \eqref{eqn:virialPressure} at different radii $r$, given here in units of the kernel support radius $H$. The actual radius of this spherical droplet is $1\mathrm{mm}$, approximately $8.5H$. The calculated virial pressure diverges for $r \gtrsim 6.5H$ due to the neighbourhood deficiency of the particles near the free surface. A red line shows the region over which we average $P(r)$.}
    \label{fig:virialPressureProfile}
\end{figure}

The first test we use will validate the surface tension of the fluid in a static scenario, independent of fluid-solid interactions. 
The Young-Laplace equation describes the difference in pressure $\Delta P$ due to surface tension $\sigma$ in a spherical droplet of radius $R$ \citep{bormashenkoPhysics2017}:
\[ 
\Delta P = \frac{2\sigma}{R}.
\]
By testing the pressure difference at different droplet radii for a fixed inter-particle force strength $s_\mathrm{ff}$, we can calibrate (and validate) the resulting surface tension $\sigma$. 
Given that we only model the fluid phase, and therefore have $P_\text{out} = 0$, we need only calculate $P_\text{in}$.
We do this by borrowing a technique from molecular dynamics, calculating the total pressure from a many-body simulation \citep{hooverIsomorphism1998}.
When calculated this way, the pressure due to particle-particle interactions is called virial pressure, and is defined by \cite{tartakovskyModeling2005,allenComputer1989} as
\begin{gather}
    P(r) = \frac{1}{2dV(r)} \sum_{i\in\mathcal{I}(r)} \sum_{j=1}^N \vec{F}_{ij} \cdot (\vec{x}_i - \vec{x}_j),
    \label{eqn:virialPressure}
\end{gather}
where $d$ is the spatial dimension (taken here to be $d=3$), $V(r)$ is the volume of a sphere of radius $r$, and $\vec{F}_{ij}$ is the sum of the pressure force and pairwise force that particle $i$ experiences due to particle $j$.
The outer summation (over $i$) includes only particles in the set $\mathcal{I}(r)$ of particles within a distance $r$ of the centre of the droplet.
This so-called virial radius $r < R$ is used in place of the actual droplet radius $R$ to exclude the region near the interface \citep{tartakovskyModeling2005}.
When $r = R$, we see a divergence of the pressure near the free surface due to the neighbourhood deficiency in the SPH approximations there.
Figure \ref{fig:virialPressureProfile} shows the virial pressure measured at different radii, with the pressure clearly diverging as $r$ approaches $R$.
For accurate estimates of the virial pressure we take an average of $P(r)$ over the interval $r \in [R-3H, R-2H]$.
If $R \leq 4H$, we take $r = R - 2H$.

\begin{figure}
    \centering
    \includegraphics[width=\textwidth]{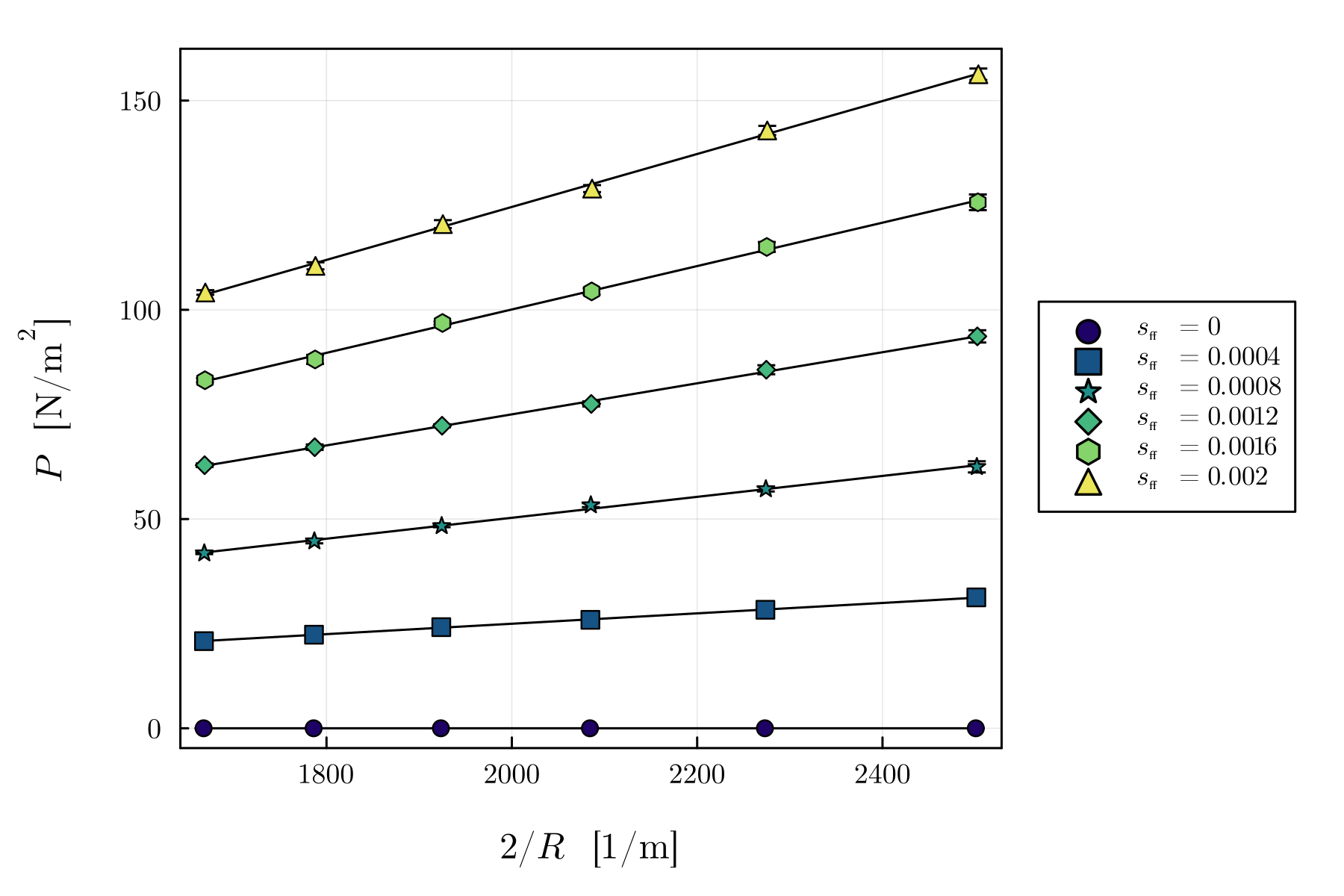}
    \caption{Laplace pressure validation using calculated virial pressures of spherical droplets. For different values of the cohesive force strength $s_\mathrm{ff}$, the pressure follows $P \propto 2 / R$, where $R$ is the radius of the droplet. Markers show measurements from simulations, and black lines show linear fits. Error bars show standard deviations of the virial pressure across the virial radii $r \in [R-3H, R-2H]$. The slope of each line gives the surface tension $\sigma$ for the corresponding $s_\mathrm{ff}$.}
    \label{fig:laplacePressure}
\end{figure}

\begin{figure}
    \centering
    \includegraphics[width=0.8\textwidth]{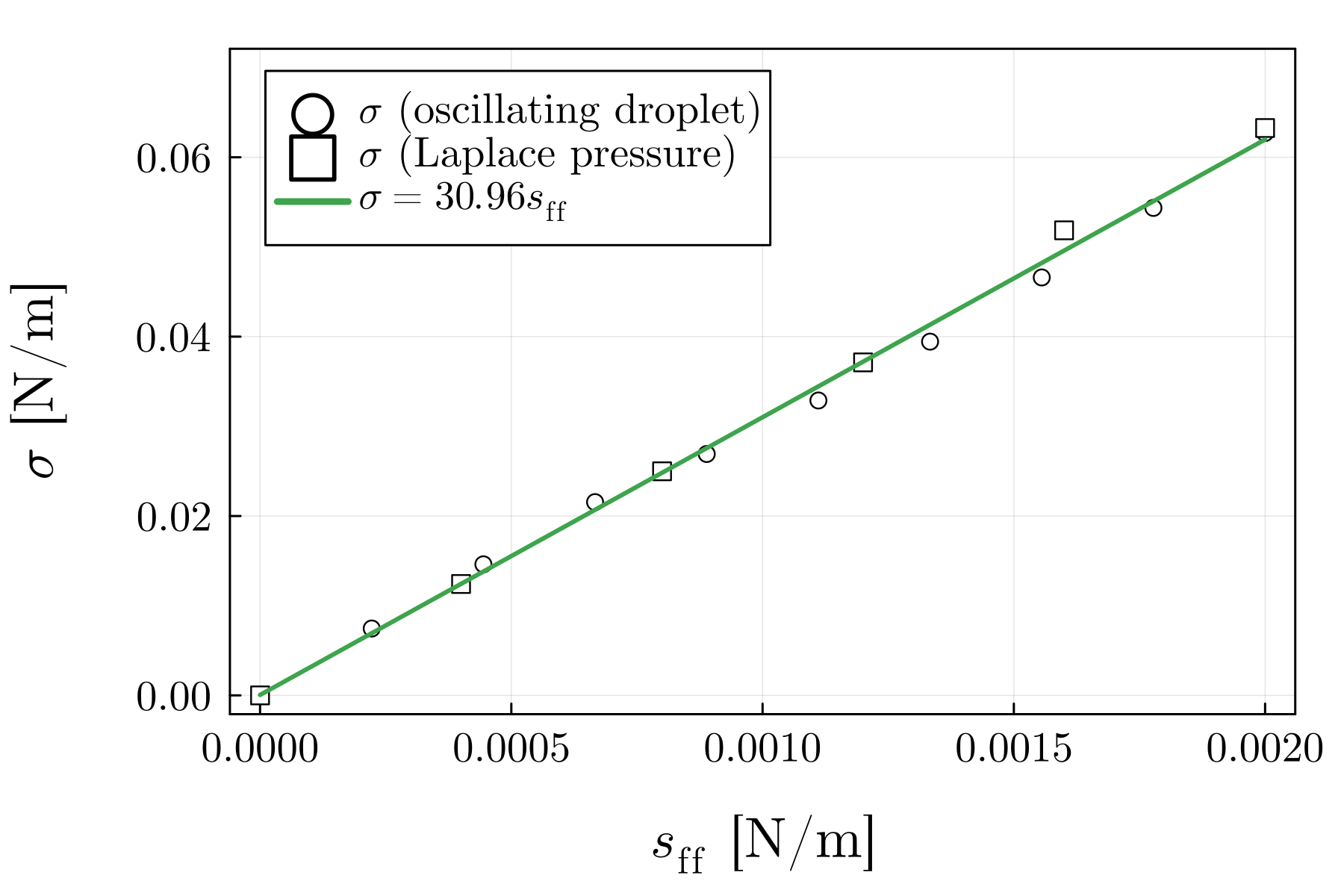}
    \caption{Calibrating the cohesive force strength $s_\mathrm{ff}$ to measured surface tension in two different tests. Circles show surface tensions calculated from the Laplace pressure $P = 2\sigma/R$ (see Figure \ref{fig:laplacePressure}). 
    Squares show surface tensions calculated from ellipsoidal droplet oscillations (e.g., Figure \ref{fig:oscillatingDiameter}).
    The fitted line shows a simple linear relationship between $s_\mathrm{ff}$ and the surface tension.}
    \label{fig:sigma_vs_sff}
\end{figure}

For the simulations, we initialise spherical droplets consisting of particles with properties given in Table \ref{tab:laplaceDropletProperties}, in zero gravity. 
We randomly perturb each particle, by no more than $0.1\Delta x$, to speed up their rearrangement due to inter-particle forces $\vec{F}^\mathrm{(pf)}$.
We then allow the droplet to reach equilibrium over $200 \,\mathrm{ms}$ before measuring the virial pressure.
Figure \ref{fig:laplacePressure} shows that the pairwise force model reproduces the linear relationship $\Delta P \propto 2/R$; we can measure the surface tension at each value of $s_\mathrm{ff}$ as the slope of each of these lines.
We also note that the standard deviation of the virial pressure over the range $r \in [ R-3H, R-2H ]$ is relatively small and thus does not introduce significant uncertainty in the calibrated surface tensions.
Plotting the measured surface tension against the model parameter $s_\mathrm{ff}$ reveals a simple linear relationship in Figure \ref{fig:sigma_vs_sff}, namely
\begin{align}
    \sigma = 30.96 s_\mathrm{ff}. 
    \label{eq:sigma_per_sff}
\end{align}
This is consistent with the dimensional analysis in Section \ref{sec:pairwiseForceModel}, and makes the prescription of surface tension in the model very simple.
We have tested this relationship for particle width $\Delta x$ as small as $2 \cdot 10^{-5} \,\mathrm{m}$ and as large as $8 \cdot 10^{-5} \,\mathrm{m}$ and found it to be independent of the resolution, as expected.

\begin{table}
    \centering
    \caption{Fluid properties in Laplace pressure simulations}
    \begin{tabular}{l l}
        \toprule
        Property & Value(s)  \\ \midrule
        Density, $\rho_0$ & $10^3 \,\mathrm{kg/m^3}$ \\
        Viscosity, $\mu$ & $0.89 \,\mathrm{mPa\cdot s}$ \\
        Speed of sound, $c_0$ & $60\,\mathrm{m/s}$ \\
        Radius, $R$ & $[0.8, 1.2] \,\mathrm{mm}$ \\
        $s_\mathrm{ff}$ & $[0, 2\cdot10^{-3}] \,\mathrm{N/m}$ \\
        Resolution, $H$ & $1.2 \cdot 10^{-4} \,\mathrm{m}$ \\
        \# Particles & $[79\,400, \, 268\,000]$ \\
        \bottomrule
    \end{tabular}
    \label{tab:laplaceDropletProperties}
\end{table}

\subsection{Oscillating droplets}
With the surface tension now calibrated in a static scenario, we next validate the model for surface tension in a dynamic scenario. 
For this task we choose to study the oscillation of a free droplet that has been perturbed from its spherical equilibrium.
The linear frequency of oscillation of an inviscid droplet (in the eigenmode of interest) was found by \cite{rayleigh1879capillary} (with a more succinct derivation given by \cite{landauFluid1987}) to be
\begin{align}
    f &= \frac{1}{\pi} \sqrt{ \frac{ 2 \sigma }{ R^3 \rho } }. 
    \label{eqn:rayleighOscillation}
\end{align}
With material properties given in Table \ref{tab:oscDropletProperties}, we initialise a spherical droplet of radius $0.788$mm with particles that we randomly perturb by no more than $0.1\Delta x$, and allow the particle distribution to settle for 1ms.
\begin{table}
    \centering
    \caption{Fluid properties in oscillating droplet simulations}
    \begin{tabular}{l l}
        \toprule
        Property & Value(s)  \\ \midrule
        Density, $\rho_0$ & $1000 \,\mathrm{kg/m^3}$ \\
        Viscosity, $\mu$ & $0$ \\
        Speed of sound, $c_0$ & $100\,\mathrm{m/s}$ \\
        Volume, $V$ & $2.05\,\mathrm{\mu L}$ \\
        $s_\mathrm{ff}$ & $[0,\, 2] \,\mathrm{mN/m}$ \\
        Resolution, $H$ & $1.2 \cdot 10^{-4} \,\mathrm{m}$ \\
        \# Particles & $75\,993$ \\
        \bottomrule
    \end{tabular}
    \label{tab:oscDropletProperties}
\end{table}
Then we `stretch' the spherical droplet into an ellipsoid with the coordinate transform
\begin{align*}
    \vec{x} \leftarrow
    \underbrace{\frac{1}{\sqrt[3]{abc}}
    \begin{bmatrix}
        a & 0 & 0 \\
        0 & b & 0 \\
        0 & 0 & c
    \end{bmatrix}}_T
    \vec{x}.
\end{align*}
Since $\det(T) = 1$, this transformation preserves volume, and therefore density of the fluid particles.
The elements $a,b,c$ are the relative lengths of the ellipsoid's semi-axes, which we take to be 1.0, 0.7, and 0.7, respectively.

\begin{figure}
    \centering
    \includegraphics[width=0.8\textwidth]{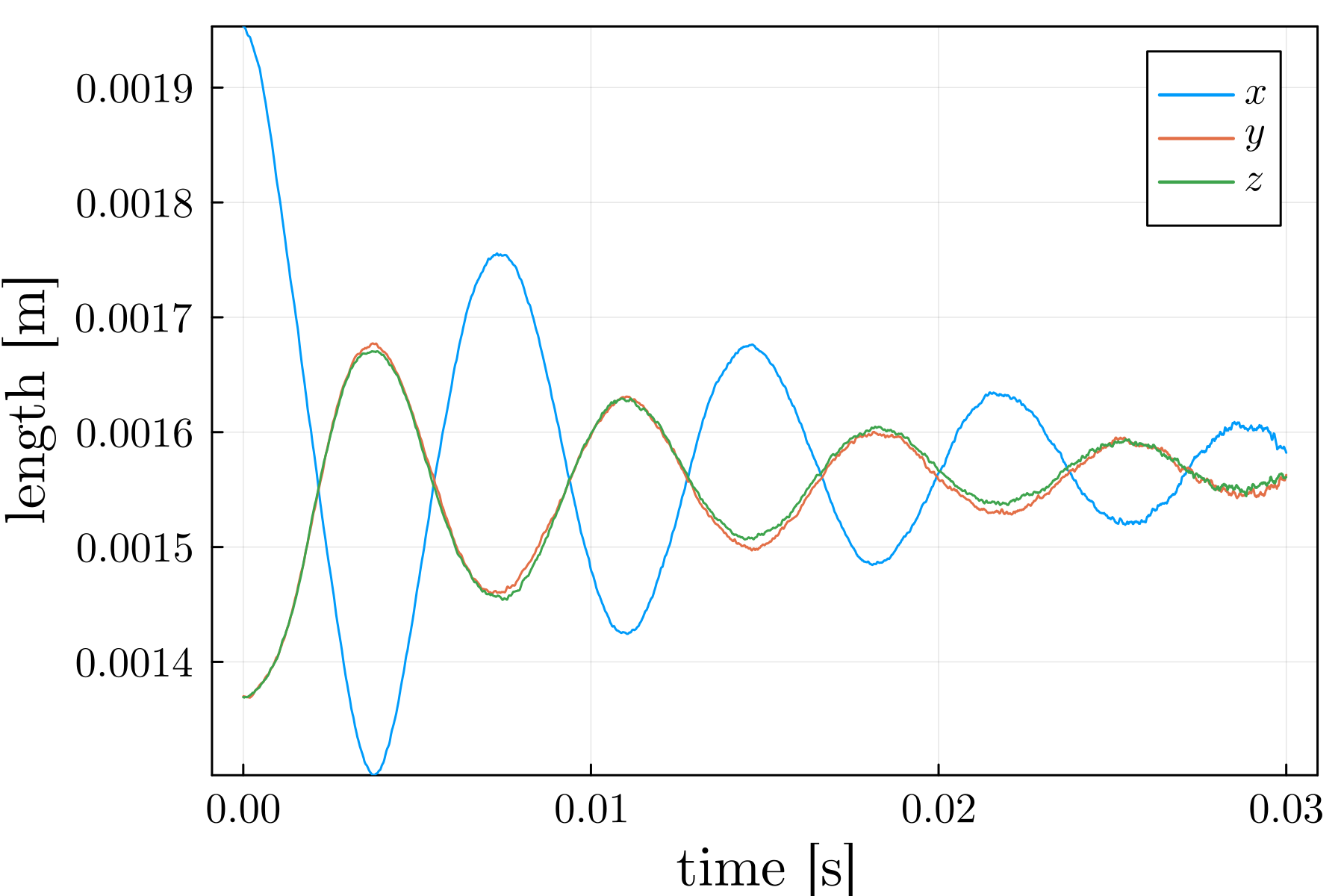}
    \caption{The oscillating diameters of an inviscid, axis-aligned ellipsoidal droplet over 30ms, for $s_\mathrm{ff} = 0.00156$. Material properties are given in Table \ref{tab:oscDropletProperties}.}
    \label{fig:oscillatingDiameter}
\end{figure}

The simulation then proceeds, with the diameter of the droplet oscillating over 3ms as shown in Figure \ref{fig:oscillatingDiameter} for $s_\mathrm{ff} = 0.00156$ (corresponding to $\sigma = 47 \,\mathrm{mN/m}$). 
Despite the fluid having zero viscosity in the simulation, the oscillations are clearly damped -- the amplitude decreases with each oscillation.
\cite{nairDynamic2018} investigate possible causes of this damping, finding that some of the system's energy is dissipated as the particles arrange themselves into a crystal-like lattice. 

\begin{figure}
    \centering
    \includegraphics[width=0.8\textwidth]{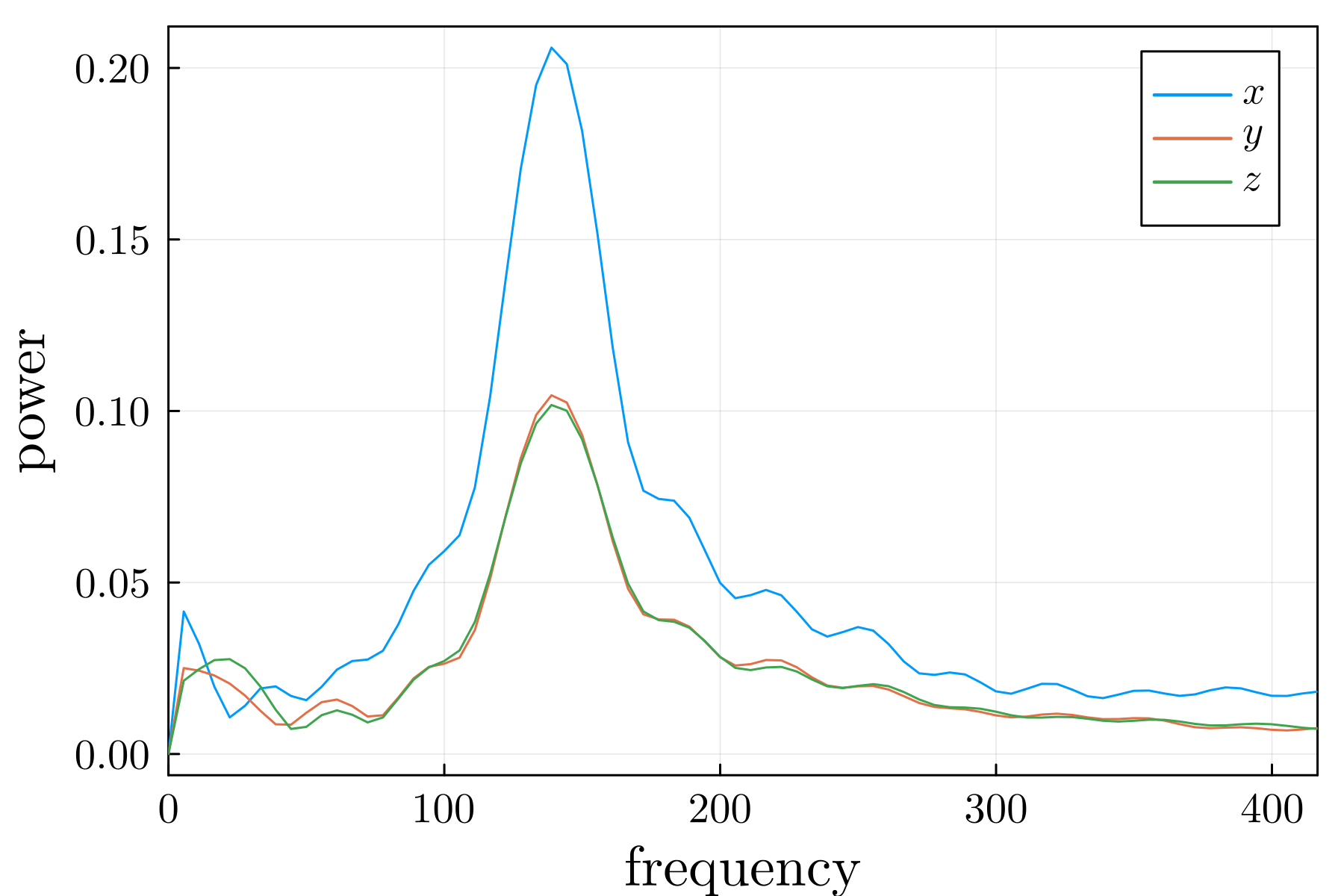}
    \caption{The power spectrum of the oscillating droplet's diameters (see Figure \ref{fig:oscillatingDiameter}), with a peak at 138Hz, which gives $\sigma = 47 \,\mathrm{mN/m}$ when $s_\mathrm{ff} = 0.00156$ from equation \eqref{eqn:rayleighOscillation}.}
    \label{fig:oscillatingFFT}
\end{figure}

Despite these effects, we can recover the frequency of the oscillations to determine the surface tension of the droplet.
A discrete Fourier transform of the oscillations (Figure \ref{fig:oscillatingFFT}) shows a peak at 138Hz. 
With equation \eqref{eqn:rayleighOscillation}, we can calculate the surface tension as $\sigma = 47 \,\mathrm{mN/m}$ when $s_\mathrm{ff} = 0.00156$.
Figure \ref{fig:sigma_vs_sff} shows the surface tensions calculated this way, plotted against the cohesive force $s_\mathrm{ff} \in [0,0.002]$, corroborating the linear relationship from the Laplace pressure test in Section \ref{sec:laplace}.
Thus, we have shown in two independent tests -- one static and the other dynamic -- that the surface tension induced by the pairwise forces scales linearly with the interaction strength parameter $s_\mathrm{ff}$.

\subsection{Young-Laplace profile fitting for contact angle}\label{sec:YLprofiles}
Having calibrated and validated the surface tension in both static and dynamic scenarios, we now turn our attention to wetting phenomena and the adhesive force strength $s_\mathrm{fs}$.
The angle that a droplet's liquid-gas interface makes with its liquid-solid interface -- the contact angle -- is commonly used to measure the ability of the liquid to wet the solid \citep{bormashenkoPhysics2017}.
This angle is often measured experimentally by drawing a tangent line to the liquid-gas interface where it meets the solid; however, this would seem only to validate the model in the immediate vicinity of the contact line. 
Instead, we will use semi-analytical solutions for whole droplet shapes to validate the model and calibrate the adhesive force to the contact angle.

For this test, we initialise a spherical droplet at a distance of $H$ (the kernel support radius) above a flat solid surface, with a thickness of $H$, to ensure the neighbourhood of a fluid particle near the boundary is not deficient. 
In all the contact angle simulations, the fluid particles have density $1000\,\mathrm{kg/m^3}$, viscosity $0.89\,\mathrm{mPa\cdot s}$, and artificial speed of sound $100\,\mathrm{m/s}$.
We have simulated droplets of various surface tensions, volumes, and resolutions to ensure the model is not scale-dependent. 
Initially, the fluid particles are perturbed by no more than $0.1 \Delta x$ and the particle distribution is allowed to settle for $1\,\mathrm{ms}$ under zero gravity, without interacting with the surface.
Then, we switch on gravity ($g = -9.81\,\mathrm{m/s^2}$), allowing the droplet to fall and spread across the solid surface.
After $50\,\mathrm{ms}$, the droplet settles and we are ready to extract the liquid-gas interface.

The liquid-gas interface of an SPH-simulated droplet is an isosurface of the density field defined by the SPH interpolation $\langle \rho(\vec{x}) \rangle$:
\begin{equation}
    \mathcal{S} = 
    \left\{ \vec{x} \,:\, \sum_{j \,\in\,\mathcal{I}_\mathrm{f}} m_j W(\vec{x} - \vec{x}_j; H) = \frac{\rho_0}{2} \right\},
    \label{eq:densityIsosurface}
\end{equation}
where $\mathcal{I}_\mathrm{f}$ is the index set of fluid particles.
We realise the implicit surface $\mathcal{S}$ using the marching cubes algorithm, with a grid spacing of $0.1\Delta x$.
Then, to determine the contact angle of the droplet, we fit the shape of an axisymmetric sessile droplet determined by the Young-Laplace (Y-L) equation \citep{danovShape2016} to the vertices of the SPH droplet isosurface in cylindrical polar coordinates $(R_j,Z_j)$ where the origin is given by the droplet's centre of mass.

In order to fit the data points to the Y-L surface we minimise the function
\[
L(\theta_\mathrm{CA},Z_\text{shift}) = 
\sum_j \frac{1}{R_j + \epsilon} \, \mathrm{mindist}(R_j,Z_j-Z_\text{shift};\theta_\mathrm{CA})^2,
\]
where $\theta_\mathrm{CA}$ is the contact angle, $Z_\text{shift}$ is a vertical shift of the data position, and $\mathrm{mindist}(R,Z;\theta_\mathrm{CA})$ is the minimum distance from the point $(R,Z)$ to the Y-L surface for the specified volume, surface tension and contact angle.
The circumference of the droplet, and therefore the number of expected isosurface vertices, increases linearly with $R$.
Thus, to account for the varying density of the samples, we weight the errors with the factor $1 / (R_j + \epsilon)$, where $\epsilon > 0$ is a small constant to avoid division by zero. 

To construct the distance function $d(R,Z;\theta_\mathrm{CA})$ we must first compute the Y-L droplet shape for a given volume $V_0$, surface tension $\sigma$, and contact angle $\theta_\mathrm{CA}$. Following \cite{danovShape2016} (using equations first reported by \citet[p. 40]{hartlandAxisymmetric1976}), we solve the system of ordinary differential equations
\begin{align}
\begin{split}
    \frac{\mathrm{d}r}{\mathrm{d}s}&=\cos\theta,\\
    \frac{\mathrm{d}z}{\mathrm{d}s}&=\sin\theta,\\
    \frac{\mathrm{d}\theta}{\mathrm{d}s}&=\frac{2}{\mathcal{B}}-\frac{\sin\theta}{r}+\frac{\rho g}{\sigma}z,\\   
    \frac{\mathrm{d}V}{\mathrm{d}s}&=r^2\sin\theta,
\end{split}
\label{eq:YL_ODEs}
\end{align}
from the ``initial condition'' $r=z=\theta=V=0$ until $\theta=\theta_\mathrm{CA}$, where $\mathcal{B}$ is the curvature at the apex of the droplet, chosen such that the correct volume is achieved. Numerically, the Y-L drop surface is given as a series of points $(r_i,z_i)$ with their gradient as an angle $\theta_i$.

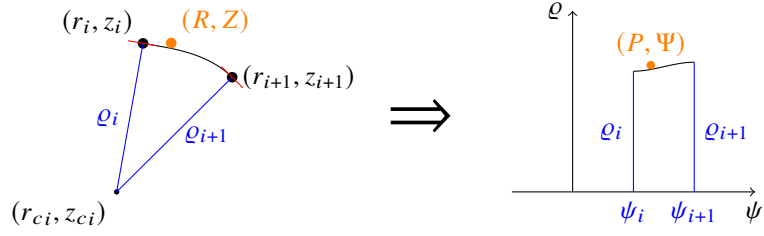
\begin{figure}
    \centering
    \begin{tikzpicture}
    \usetikzlibrary{calc}
    \def\rhoOne{2}
    \def\rhoTwo{2.15}
    \def\rhoX{2.1}
        \coordinate (o) at (0,0);
        \coordinate (p1) at ({\rhoOne*cos(80)},{\rhoOne*sin(80)});
        \coordinate (p2) at ({\rhoTwo*cos(45)},{\rhoTwo*sin(45)});
        \coordinate (px) at ({\rhoX*cos(70)},{\rhoX*sin(70)});
        \draw[domain=0:1, smooth, variable=\t] plot ({(\rhoOne+(\rhoTwo-\rhoOne)*(3*\t^2-2*\t^3))*cos(80-35*\t)}, {(\rhoOne+(\rhoTwo-\rhoOne)*(3*\t^2-2*\t^3))*sin(80-35*\t)});
        \fill[black]  (o) circle (1pt) node[below left] {$({r_c}_i,{z_c}_i)$};
        \fill[orange]  (px) circle (2pt) node[above right] {$(R,Z)$};
        \fill[black]  (p1) circle (2pt) node[above left] {$(r_i,z_i)$};
        \fill[black]  (p2) circle (2pt) node[right] {$(r_{i+1},z_{i+1})$};
        \draw[blue] (o) -- (p1);
        \node[blue,left] at ($(o)!0.5!(p1)$) {$\varrho_i$};
        \draw[blue] (o) -- (p2);
        \node[blue,right] at ($(o)!0.5!(p2)$) {$\varrho_{i+1}$};
        \draw[red] (p1) -- +({0.2*cos(-10)},{0.2*sin(-10)});
        \draw[red] (p1) -- +({-0.2*cos(-10)},{-0.2*sin(-10)});
        \draw[red] (p2) -- +({0.2*cos(-45)},{0.2*sin(-45)});
        \draw[red] (p2) -- +({-0.2*cos(-45)},{-0.2*sin(-45)});
        \node at (4,1) {\Huge $\Rightarrow$};
        \begin{scope}[xshift=6cm,scale=0.8]
            \draw[->] (-1,0) -- (3,0) node[below] {$\psi$};
            \draw[->] (0,0) -- (0,3) node[left] {$\varrho$};
            \draw[blue] (1,0) node[below] {$\psi_i$} -- (1,\rhoOne);
            \node[blue,left] at (1,1) {$\varrho_i$};
            \draw[blue] (2,0) node[below] {$\psi_{i+1}$} -- (2,\rhoTwo);
            \node[blue,right] at (2,1) {$\varrho_{i+1}$};
            \draw[domain=0:1, smooth, variable=\t] plot ({\t+1}, {\rhoOne+(\rhoTwo-\rhoOne)*(3*\t^2-2*\t^3)});
            \fill[orange]  ({1+10/35},{\rhoX}) circle (2pt) node[above] {$(P,\Psi)$};
        \end{scope}
    \end{tikzpicture}
    \caption{Schematic of the transformation from the cylindrical polar world coordinates to the local polar coordinates of a surface segment.}
    \label{fig:normal-distance-transform}
\end{figure}

We approximate the normal distance to the surface by transforming the coordinates for a given surface segment $(r_i,z_i)-(r_{i+1},z_{i+1})$ into a local polar coordinate system 
\[
(\varrho,\psi)\rightarrow\left(\sqrt{(r_i-r_c)^2+(z_i-z_c)^2},\arctan\frac{z_i-z_c}{r_i-r_c}\right),
\] 
where the centre $(r_c,z_c)$ is given by the intersections of two lines passing through each of the end points of the segment perpendicular to their gradients (Figure \ref{fig:normal-distance-transform}). That is,
\begin{align*}
    r_c=&\frac{\sin\theta_i(z_{i+1}-z_i)+\cos\theta_i(r_{i+1}-r_i)}{\sin\theta_i\cos\theta_{i+1}-\cos\theta_i\sin\theta_{i+1}}\sin\theta_i+r_i,\\
    z_c=&-\frac{\sin\theta_i(z_{i+1}-z_i)+\cos\theta_i(r_{i+1}-r_i)}{\sin\theta_i\cos\theta_{i+1}-\cos\theta_i\sin\theta_{i+1}}\cos\theta_i+z_i.
\end{align*}
The surface location along the segment is then given by $\varrho_i(\psi)=\varrho_i+(\varrho_{i+1}-\varrho_i)(3t^2-2t^3)$, where $t=(\psi-\psi_i)/(\psi_{i+1}-\psi_i)$. The difference $P-\varrho_i(\Psi)$ is taken as the approximation of the normal distance from the point $(R,Z)$ to the segment. Therefore the minimum distance from the point $(R,Z)$ to the surface is the minimum distance over all segments for which the surface interpolation is valid,
\[
\mathrm{mindist}(R,Z;\theta_\mathrm{CA}) = \min_i \min_{\Psi\in[\psi_i,\psi_{i+1}]}\left|P-\varrho_i(\Psi)\right|.
\]

\begin{figure}
    \centering
    \includegraphics[width=0.8\textwidth]{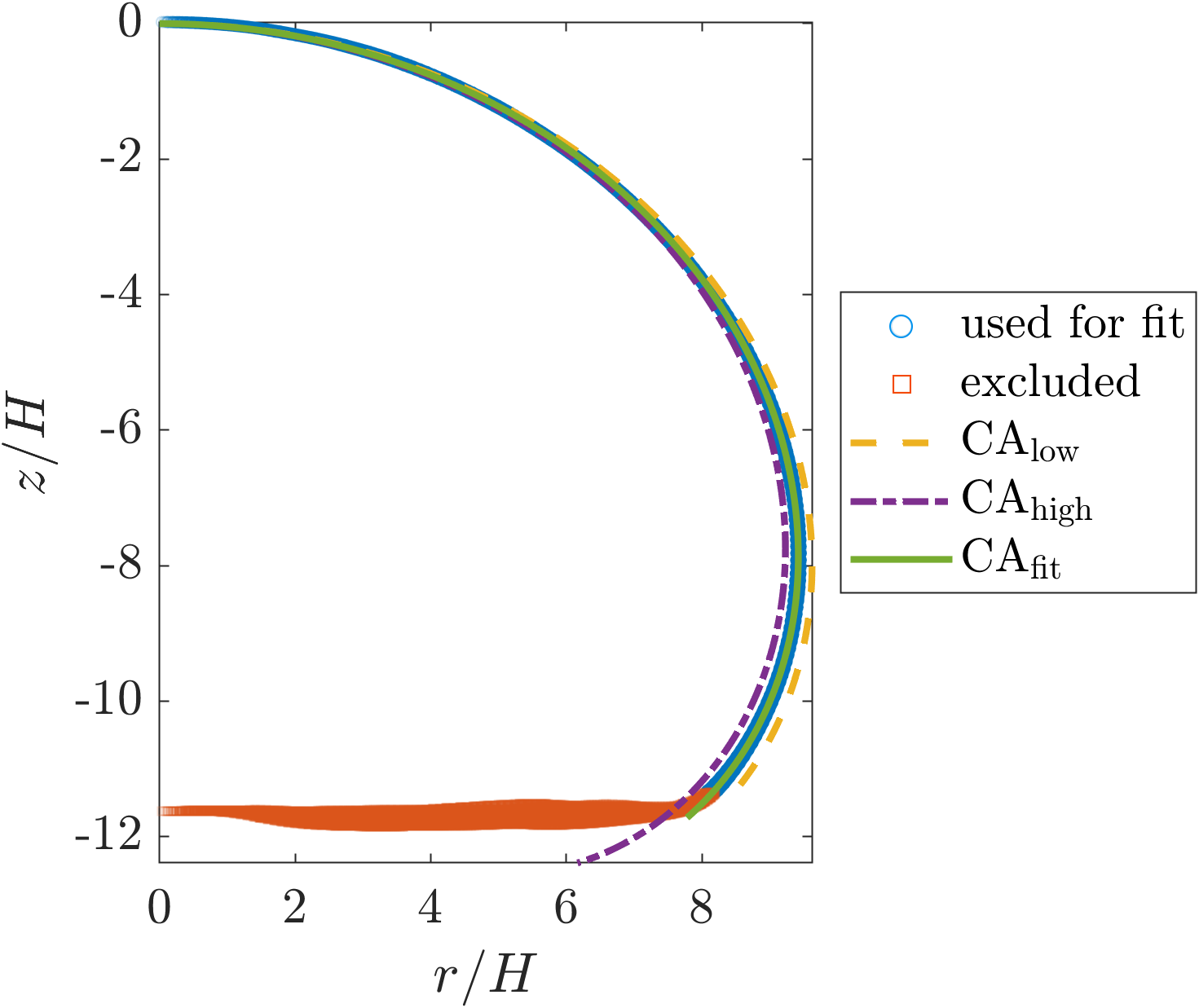}
    \caption{An example of a Young-Laplace profile (a solution to the system \eqref{eq:YL_ODEs}) fitted to an SPH droplet. Isosurface vertices are shown in cylindrical coordinates about the centre of the top of droplet. The solid line shows the profile of an axis-symmetric droplet satisfying the Young-Laplace equation, fit to the isosurface vertices with the contact angle as a free parameter. Dashed and dot-dashed lines show the extent of contact angles satisfying the fitting tolerance.}
    \label{fig:CAcalExample}
\end{figure}

\begin{figure}
    \centering
    \includegraphics[width=\textwidth]{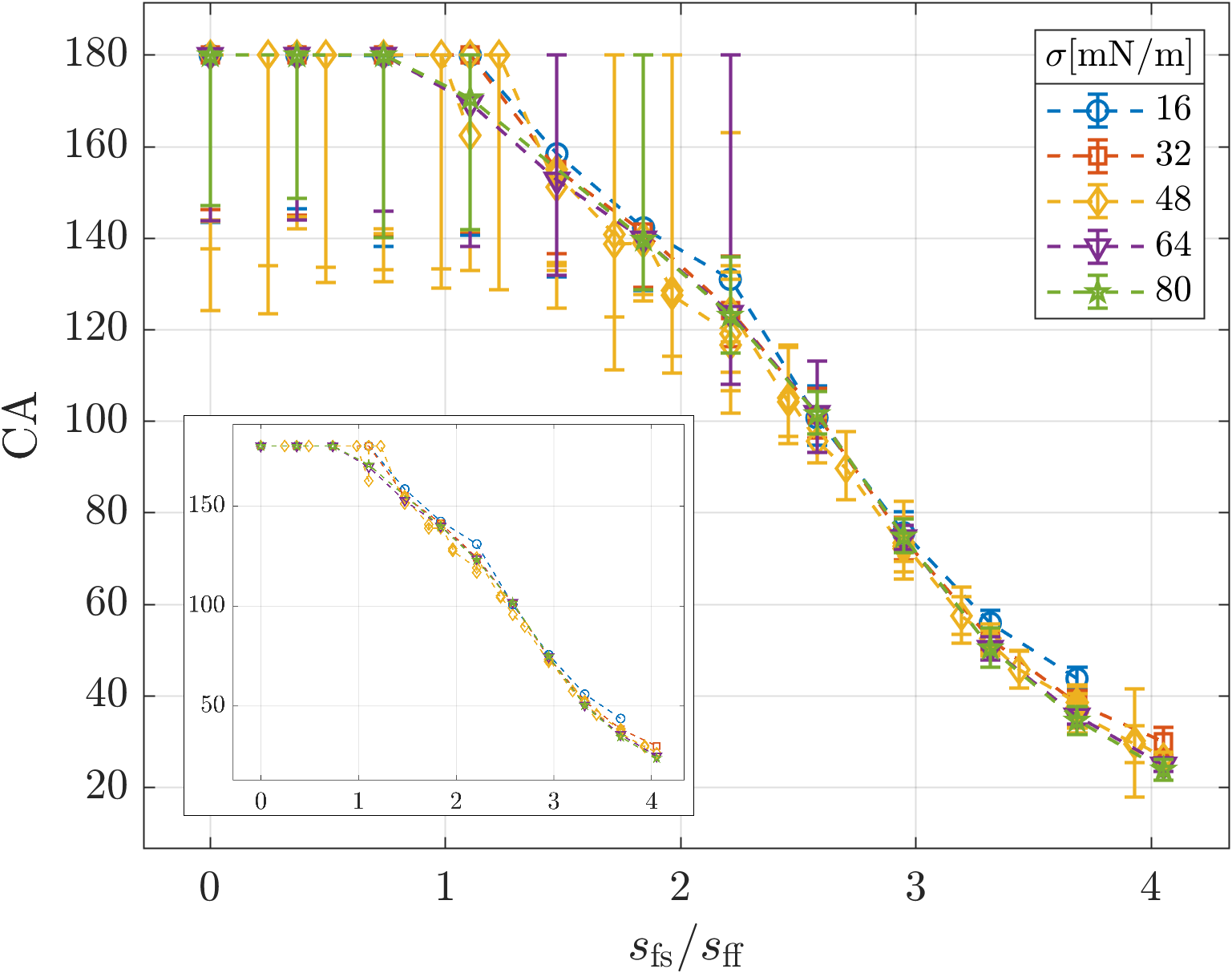}
    \caption{Contact angles, measured by fitting Young-Laplace profiles to sessile droplet density isosurfaces, are plotted against the ratio of pairwise force strengths, $s_\mathrm{fs}/s_\mathrm{ff}$, and grouped by surface tension. A droplet is excluded if the volume enclosed by the density isosurface differs from the actual droplet's volume by more than 10\%. Inset: the curves without error bars, for clarity.}
    \label{fig:CA_vs_sratio}
\end{figure}

\begin{figure}
    \centering
    \includegraphics[width=\textwidth]{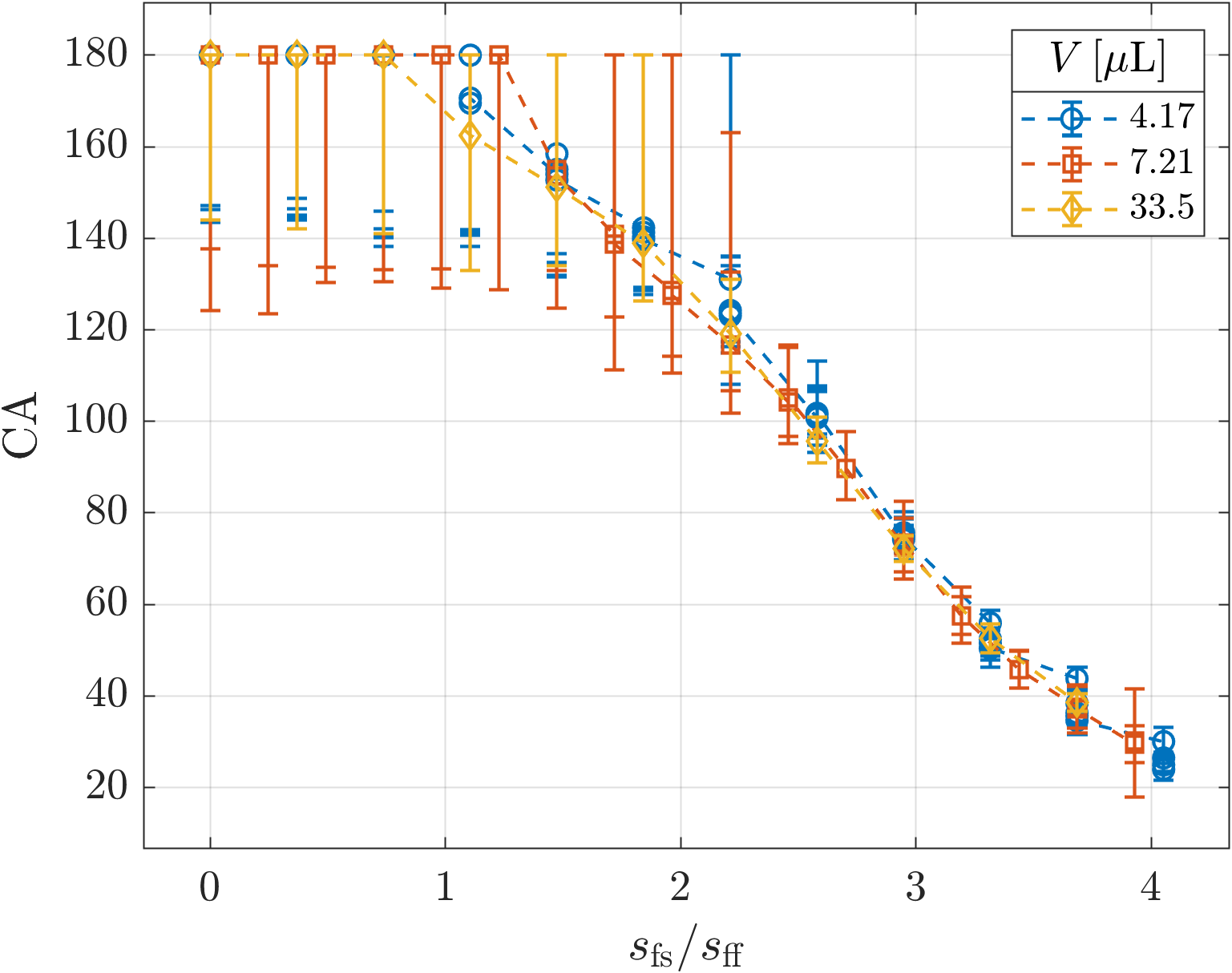}
    \caption{Contact angles, measured by fitting Young-Laplace profiles to sessile droplet density isosurfaces, are plotted against the ratio of pairwise force strengths, $s_\mathrm{fs}/s_\mathrm{ff}$, and grouped by volume. A droplet is excluded if the volume enclosed by the density isosurface differs from the actual droplet's volume by more than 10\%.}
    \label{fig:CA_vs_sratio_volume}
\end{figure}

Figure \ref{fig:CAcalExample} shows an example of an SPH droplet isosurface and its best fit Y-L shape. 
Also shown are the Y-L shapes for the contact angles $\theta_\text{low}$ and $\theta_\text{high}$, such that $\theta_\text{CA} \in [\theta_\text{low}, \theta_\text{high}]$ and $L(\theta_\text{low}, Z_\text{shift}) = L(\theta_\text{high}, Z_\text{shift}) = \texttt{tol}$, a fitting tolerance.
For each simulated droplet, this approach gives a feasible range of contact angles as well as the optimal fit.
The results of measuring the contact angles of simulated droplets in this way are shown in Figures \ref{fig:CA_vs_sratio} and \ref{fig:CA_vs_sratio_volume}, grouped by surface tension and volume, respectively, and plotted against the ratio $s_\mathrm{fs} / s_\mathrm{ff}$. 
The contact angles for different surface tensions and volumes coincide for equal values of this ratio, suggesting that the contact angle produced by the PF-SPH model is scale-independent and making the specification of the contact angle straightforward in practice. 
We find that for higher contact angles, where larger changes in $\theta_\text{CA}$ result in only small changes in the droplet profile, the feasible range $\theta_\text{high} - \theta_\text{low}$ is relatively large.
For intermediate contact angles between $30^\circ$ and $120^\circ$, however, the feasible region is smaller and we can be more precise in our specification of $\theta_\text{CA}$ via $s_\mathrm{fs}$.

With this validation of the droplet shape and calibration of the contact angle, we now have a simple procedure for specifying the surface tension $\sigma$ and contact angle $\theta_\text{CA}$ in the PF-SPH scheme. 
We can calculate the required cohesive strength $s_\mathrm{ff} = \sigma / 30.96$ from equation \eqref{eq:sigma_per_sff}.
Then we can interpolate the data from Figure \ref{fig:CA_vs_sratio} to find the required ratio $s_\mathrm{fs}/s_\mathrm{ff}$, and, knowing $s_\mathrm{ff}$, calculate $s_\mathrm{fs}$.

%% file: src/4results.tex
In this section we demonstrate the versatility of our PF-SPH scheme by presenting some numerical experiments that are usually challenging for competing computational frameworks such as interface-tracking or grid-based methods.
Of course, these problems are not intractable for such methods -- see, for example, works such as \cite{jansenSimulating2012} in which an interface tracking method was used with a substrate with variable wettability, or \cite{dupuisModeling2005} where a Lattice-Boltzmann simulation was used with a rough substrate.
The advantage of our approach that we shall highlight is that PF-SPH is capable of modelling droplets on chemically and physically heterogeneous substrates with little to no modifications to the scheme.

\subsection{Microstructured surface}
\begin{figure}
    \centering
    \includegraphics[width=0.5\linewidth]{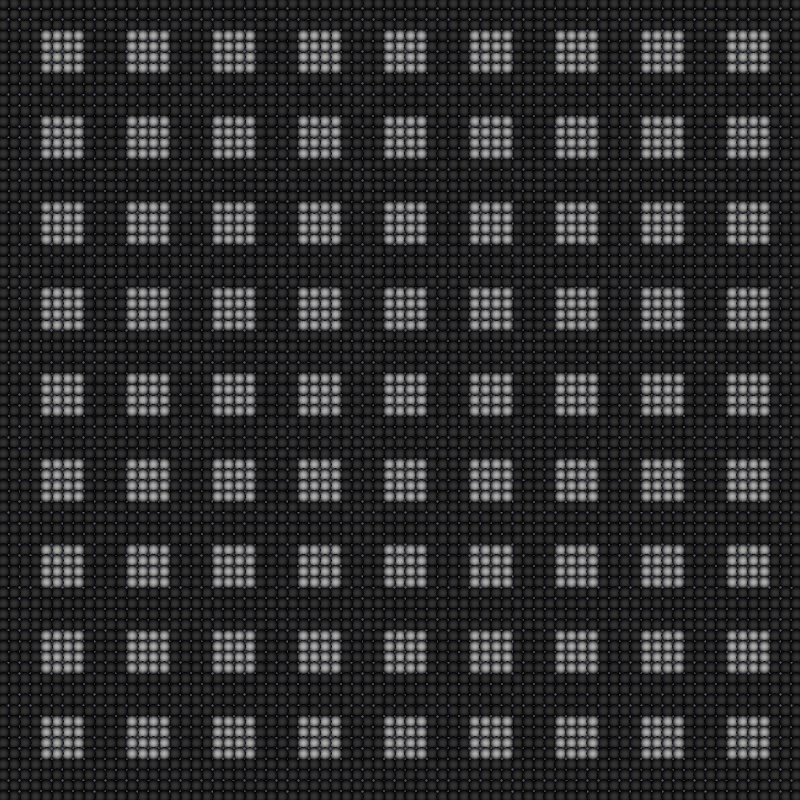}
    \caption{A top-down view of the pillared substrate used for the simulation in Figure \ref{fig:pillared}. The square pillars (shown in light grey) are $150\,\mathrm{\mu m}$ tall, $120 \,\mathrm{\mu m}$ wide, and located at regular intervals of $240\,\mathrm{\mu m}$ across the substrate (dark grey).}
    \label{fig:pillaredLayout}
\end{figure}

Synthetic surfaces with manufactured microscopic roughness attract interest from scientists and engineers alike for their potential commercial applications (e.g., self-cleaning surfaces, reduced drag on marine vessels, collecting freshwater from fog \citep{chamakosProgress2021}). 
Surfaces with sharp geometric features, however, are challenging to incorporate into computational models with explicit boundary conditions on the contact line. 
Here we will demonstrate the ease with which the calibrated PF-SPH model can be used to calculate the shape of a sessile droplet on a geometrically patterned surface.
To do this, we will follow the experimental setup of \cite{dupuisModeling2005} in simulating a $3 \,\mathrm{\mu L}$ droplet on a surface featuring square pillars, as shown in Figure \ref{fig:pillaredLayout}.
The parameters used in these simulations are given in Table \ref{tab:pillarDropletProperties}.

\begin{figure}
    \centering
    \begin{subfigure}[t]{0.48\linewidth}
        \centering
        \includegraphics[trim={0 3cm 0 1cm}, clip, width=\linewidth]{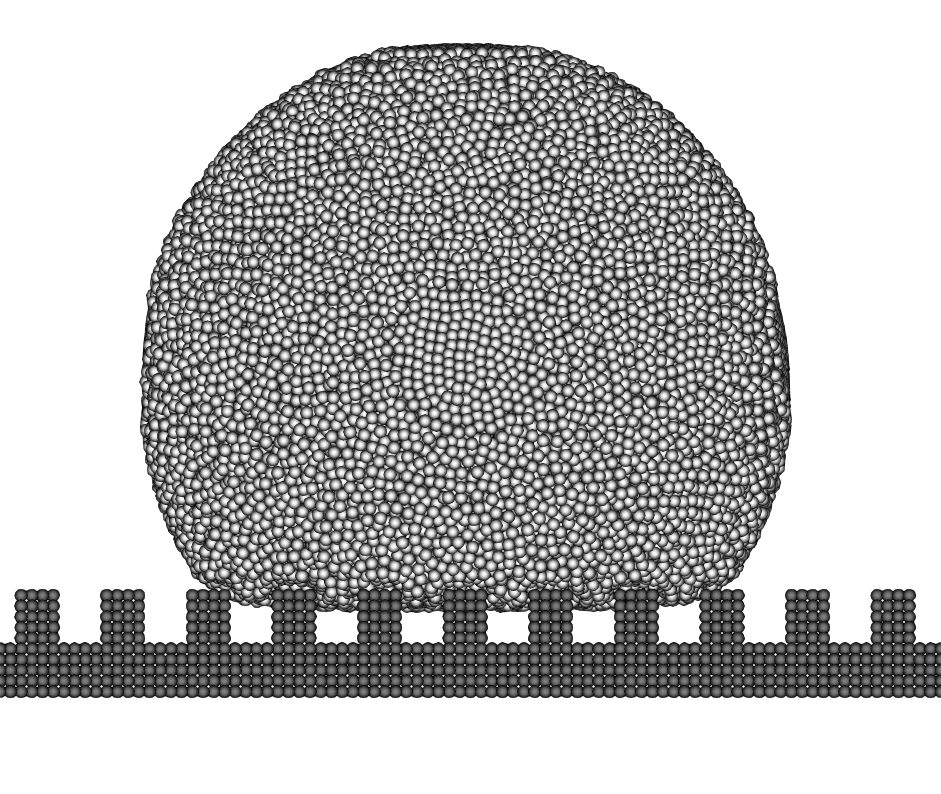}
        \caption{After 10ms of zero gravity, then 30ms under 1g, this droplet has settled, suspended atop the pillars.}
        \label{fig:pillaredGentle}
    \end{subfigure}
    \hfill
    \begin{subfigure}[t]{0.48\linewidth}
        \centering
        \includegraphics[trim={0 3cm 0 1cm}, clip, width=\linewidth]{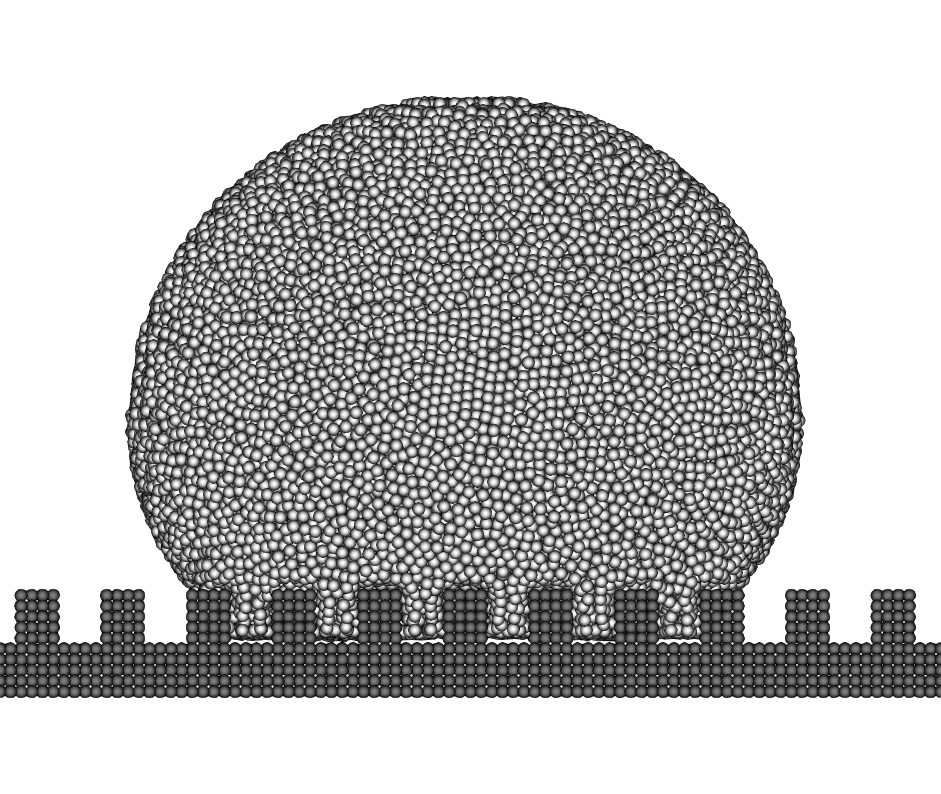}
        \caption{30ms after a 0.1m/s impact, this droplet has infiltrated the pillars.}
        \label{fig:pillaredImpact}
    \end{subfigure}
    \caption{A comparison of nearly identical simulations of a droplet settling on the physically patterned surface of square pillars depicted in Figure \ref{fig:pillaredLayout}. The addition of a small impact velocity changes the wetting behaviour significantly. (Note that these are side views of three-dimensional simulations.)}
    \label{fig:pillared}
\end{figure}

Figure \ref{fig:pillared} shows a comparison between two almost identical numerical experiments -- the only difference being the initial kinetic energy of the droplet before `impact' with the surface.
Figure \ref{fig:pillaredGentle} shows a settled droplet that was initialised with zero velocity. 
This droplet sits atop the pillars on the substrate, without enough energy to infiltrate the gaps between them.
The droplet in Figure \ref{fig:pillaredImpact} was initialised with a velocity in the vertical direction of $-0.1\,\mathrm{m/s}$, allowing it to overcome the energy barrier discussed by \cite{dupuisModeling2005} and transition from a `suspended' to a `collapsed' state, in which the fluid has infiltrated between the pillars. 
That the present model reproduces this qualitative behaviour suggests it should be applicable to problems with complex surfaces.

\begin{table}
    \centering
    \caption{Fluid properties in pillared substrate simulation in Figure \ref{fig:pillared}.}
    \begin{tabular}{l l}
        \toprule
        Property & Value(s)  \\ \midrule
        Density, $\rho_0$ & $1000 \,\mathrm{kg/m^3}$ \\
        Viscosity, $\mu$ & $0.89 \,\mathrm{mPa\cdot s}$ \\
        Speed of sound, $c_0$ & $80\,\mathrm{m/s}$ \\
        Volume, $V$ & $3.0 \,\mathrm{\mu L}$ \\
        $s_\mathrm{ff}$ & $2 \,\mathrm{mN/m}$ \\
        $s_\mathrm{fs}$ & $4.9 \,\mathrm{mN/m}$ \\
        Resolution, $H$ & $1.2 \cdot 10^{-4} \,\mathrm{m}$ \\
        \# Particles & $113\,000$ fluid particles \\
        \bottomrule
    \end{tabular}
    \label{tab:pillarDropletProperties}
\end{table}

\subsection{Chemically patterned surface}
Another type of surface heterogeneity that is widely studied is a chemically patterned surface. 
By manufacturing a surface with, for example, alternating hydrophobic and hydrophilic stripes, one can influence the shape or motion of a droplet.
\cite{varagnoloTuning2014} note that controlling the motion of very small droplets in this way is one of the key problems in the design of reliable microfluidic devices.
We will demonstrate that our PF-SPH is also a suitable model for the shape of a droplet settled on such a patterned surface.
To include the variable wettability in the PF-SPH model, we simply modify the adhesive force strength $s_\mathrm{fs}$ across the surface, depending on whether a boundary particle is hydrophobic (low wettability, high $\theta_\text{CA}$) or hydrophilic (high wettability, low $\theta_\text{CA}$).
The fluid properties for this simulation are given in Table \ref{tab:patternedDropletProperties}.
For the hydrophobic surface type, we used $s_\mathrm{fs}/s_\mathrm{ff}=1.4$ for an equilibrium contact angle of $\theta_\mathrm{hydrophobic}\approx155^\circ$, and for the hydrophilic surface type, $s_\mathrm{fs}/s_\mathrm{ff}=3.5$ for $\theta_\mathrm{hydrophilic}\approx45^\circ$.

\begin{figure}
    \centering
    \includegraphics[width=\linewidth]{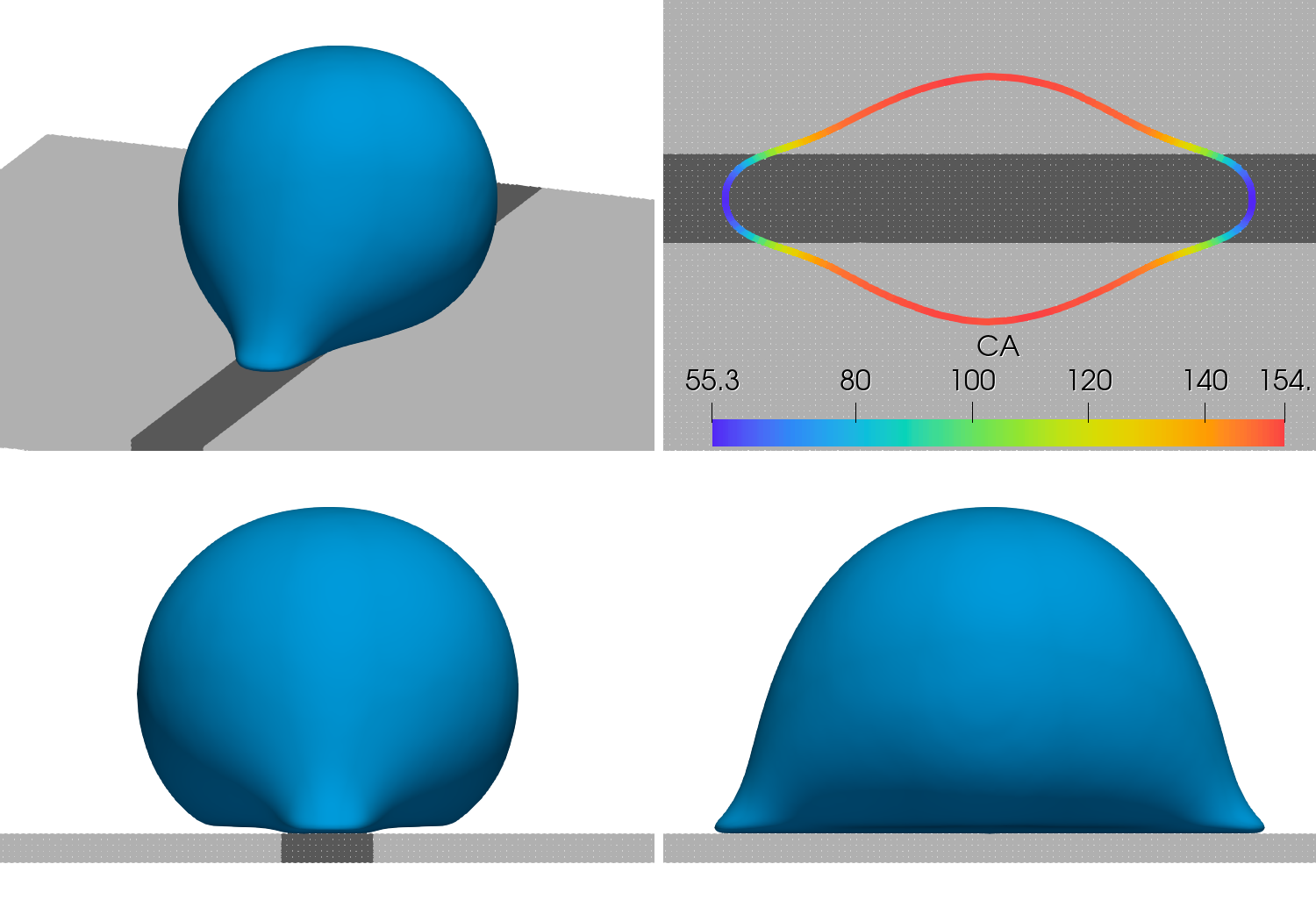}
    \caption{A droplet settles on a flat hydrophobic surface ($\theta_\mathrm{CA} \approx 155^\circ$) with a hydrophilic stripe ($\theta_\mathrm{CA} \approx 45^\circ$). Top left: isometric view. Top right: the contact angle is plotted around the contact line, showing a transition from one equilibrium contact angle to another. Bottom row: side views of the droplet, showing the significant difference in apparent contact angles on the different surface types. Surface representations of the droplet are obtained from the density isosurface; equation \eqref{eq:densityIsosurface}.}
    \label{fig:patternedDroplet}
\end{figure}

One of the advantages of our PF-SPH scheme is that the relationship between the contact angle and the position and motion of the contact line is not explicitly prescribed; instead, the simulated contact angle is ultimately a function of the adhesive force strength $s_\mathrm{fs}/s_\mathrm{ff}$ and the geometry of the substrate.
Thus, we may use the model to study the transition between contact angles on the hydrophobic and hydrophilic zones.
To visualise the contact angle on the contact line, we use the surface normal of the density isosurface \eqref{eq:densityIsosurface} at a distance of $H/2$ above the substrate.
Figure \ref{fig:patternedDroplet} shows side views of the droplet and the preferential spreading on the hydrophilic stripe. 
The contact line is shown, coloured by the contact angle, with smooth transitions between the contact angles $\theta_\mathrm{hydrophobic}$ and $\theta_\mathrm{hydrophilic}$.
The side views reveal that the contact angle the droplet makes is vastly different on the hydrophilic substrate sections as compared to the hydrophobic sections.
At its extremities, the measured contact angle reaches the specified hydrophobic value of $155^\circ$, whereas the specified hydrophilic contact angle of $45^\circ$ is not realised in the simulation, with the lowest measured contact angle being $55^\circ$. 
That our model is capable of producing such behaviour suggests it could be used to design and study manufactured, variable-wettability substrates for droplet motion control.

\begin{table}
    \centering
    \caption{Fluid properties in patterned substrate simulation in Figure \ref{fig:patternedDroplet}.}
    \begin{tabular}{l l}
        \toprule
        Property & Value(s)  \\ \midrule
        Density, $\rho_0$ & $1261 \,\mathrm{kg/m^3}$ \\
        Viscosity, $\mu$ & $5.9 \,\mathrm{mPa\cdot s}$ \\
        Speed of sound, $c_0$ & $80\,\mathrm{m/s}$ \\
        Volume, $V$ & $0.165 \,\mathrm{\mu L}$ \\
        $s_\mathrm{ff}$ & $2 \,\mathrm{mN/m}$ \\
        $s_\mathrm{fs}$ & $7 \text{ (hydrophilic)},\; 2.8 \text{ (hydrophobic)} \,\mathrm{mN/m}$ \\
        Resolution, $H$ & $4 \cdot 10^{-5} \,\mathrm{m}$ \\
        \# Particles & $164\,968$ fluid particles \\
        \bottomrule
    \end{tabular}
    \label{tab:patternedDropletProperties}
\end{table}

\subsection{Wheat leaf}
The broader context and motivation of this work is to enable the study of droplets impacting and settling on plant leaves, for which a key challenge is calculating the shape of a droplet on a plant leaf with microscopic roughness, and chemical heterogeneity \citep{mayoSimulating2015}.
Here, we will demonstrate the model's applicability to complex surfaces -- specifically, a reconstructed wheat leaf surface.
This virtual wheat leaf was reconstructed from a microCT scan using a radial basis function partition of unity method as described in related work: \cite{whebellData2025}. 
We discretise this surface for an SPH simulation by taking a block of boundary particles on a regular grid, and discarding any for which the implicit surface indicator function is negative ($\mathcal{F}(\vec{x}) < 0$). 
Manually selected boundary particles are labelled as hairs and assigned a higher adhesive force strength to reflect the surface chemistry of real wheat leaves.
The parameters for this simulation are given in Table \ref{tab:wheatDropletProperties}.

\begin{table}
    \centering
    \caption{Fluid properties in wheat leaf simulation in Figure \ref{fig:wheat}.}
    \begin{tabular}{l l}
        \toprule
        Property & Value(s)  \\ \midrule
        Density, $\rho_0$ & $1000 \,\mathrm{kg/m^3}$ \\
        Viscosity, $\mu$ & $0.89 \,\mathrm{mPa\cdot s}$ \\
        Speed of sound, $c_0$ & $100\,\mathrm{m/s}$ \\
        Volume, $V$ & $0.27 \,\mathrm{\mu L}$ \\
        $s_\mathrm{ff}$ & $2 \,\mathrm{mN/m}$ \\
        $s_\mathrm{fs}$ & $5.6 \text{ (leaf)}, \; 7.875 \text{ (hairs)} \,\mathrm{mN/m}$ \\
        Resolution, $H$ & $4 \cdot 10^{-5} \,\mathrm{m}$ \\
        \# Particles & $267\,731$ fluid particles \\
        \bottomrule
    \end{tabular}
    \label{tab:wheatDropletProperties}
\end{table}

\begin{figure}
    \centering
    \includegraphics[width=\linewidth]{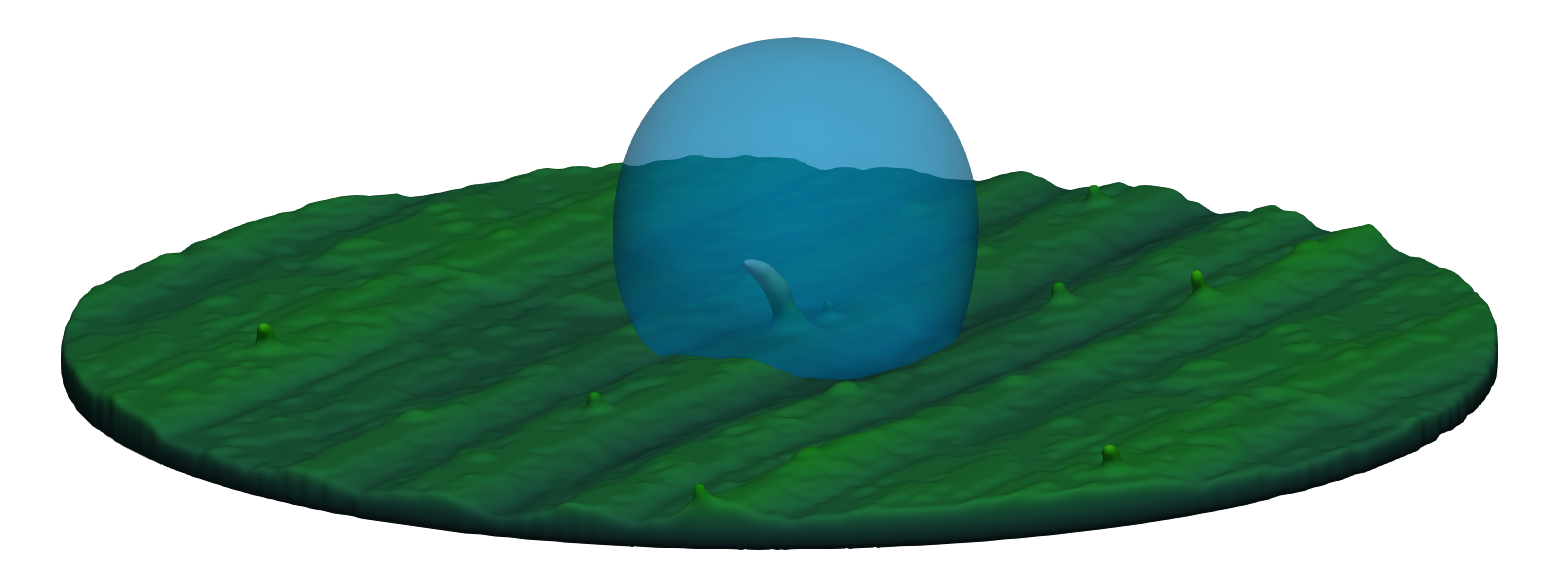}
    \caption{A pairwise force SPH simulation of a $0.27\mathrm{\mu L}$ water droplet settled on the surface of a wheat leaf after $12\,\mathrm{ms}$. The droplet's liquid-gas interface is rendered by polygonising an isosurface of the density field, $\langle\rho(\vec{x})\rangle = \rho_0 / 2$. The leaf surface is rendered similarly, using only boundary particles, and coloured by height for clarity. A hair on the leaf surface is visible under the translucent droplet.}
    \label{fig:wheat}
\end{figure}

\begin{figure}
    \centering
    \includegraphics[width=\linewidth]{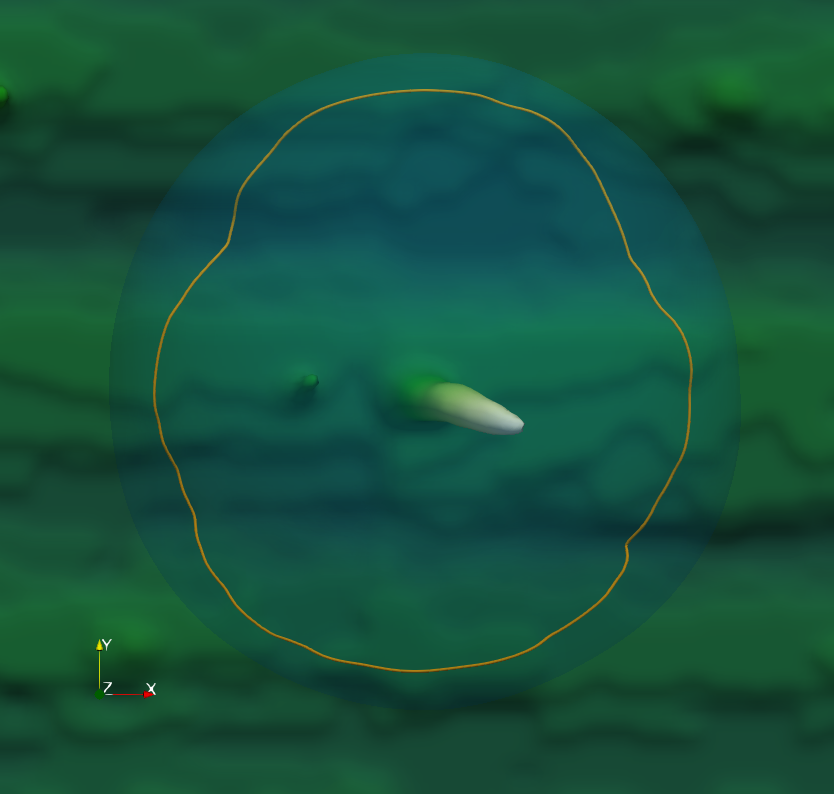}
    \caption{The contact line of the droplet on a wheat leaf depicted in Figure \ref{fig:wheat}, visualised by computing the set of points on the density isosurface that are $2\Delta x$ from the surface.}
    \label{fig:wheat_contactLine}
\end{figure}

We initialised the simulation with a sphere of fluid particles just above a hair of the wheat leaf, with zero impact velocity, to study how the droplet settles onto the surface. 
Figure \ref{fig:wheat} shows the shape of the droplet at 50ms, once it has lost momentum and settled on the leaf.
We can visualise the contact line with a contour line on the density isosurface, where the approximate minimum distance to the leaf surface is $2\Delta x$.
Specifically, Figure \ref{fig:wheat_contactLine} shows the contact line visualised as the set of points
\[
    \left\{ \vec{x} \,:\, \sum_{j \,\in\,\mathcal{I}_\mathrm{f}} m_j W(\vec{x} - \vec{x}_j; H) = \frac{\rho_0}{2} \right\}
    \cap
    \left\{ \vec{x} \,:\, \min_{\vec{y} \in \mathcal{L}} \eucnorm{\vec{x} - \vec{y}} = 2 \Delta x \right\},
\]
where $\mathcal{L}$ is the set of leaf surface points.
Note the highly irregular contact line following the natural curvature of the wheat leaf.
This droplet shape could be used to measure the wetted area of the leaf surface or serve as an initial condition for an evaporation model in future work.

\subsection{Discussion on pairwise force profiles} 
\label{sec:pf_profiles}
The exact design of the pairwise force profile $f_{ij}$ is in need of more investigation before the PF-SPH model can be applied as readily as, for example, the continuum surface force formulation.
Literature on the effect of the force profile is scarce. 
\cite{tartakovskyPairwise2016} report some analytical results predicting the surface tension from $f_{ij}(\tau)$ and $s_{ij}$ in multiphase simulations, but we could not verify these results in our single phase simulations. 
Our force profile causes most fluid-fluid particle interactions to be attractive, which could cause excess numerical stress in the fluid. 
In the simulations tested, the mostly-attractive pairwise force has the somewhat desirable effect of keeping the particle distribution ordered -- for example, in a high velocity impact event -- which ensures that interpolation error is low. 
More investigation is needed to quantify the dissipative effect of the pairwise forces before this model is applied to viscosity-dependent scenarios.

%% file: src/5conclusion.tex
We have presented a weakly compressible pairwise-force smoothed particle hydrodynamics model and applied it to study droplet shapes on complex surfaces. 
A new physically motivated pairwise force profile $f_{ij}(\tau)$ has been validated and calibrated to relate cohesive and adhesive parameters $s_\mathrm{ff}$ and $s_\mathrm{fs}$ to physical values of the surface tension and contact angle. 
Furthermore, we have described a method for calibrating the contact angle of PF-SPH models and shown that the liquid-gas interfaces of simulated droplets on flat surfaces are in good agreement with semi-analytical solutions to the Young-Laplace equations for a range of contact angles between $40^\circ$ and $180^\circ$. 
The pairwise force model is scale-independent and, since it does not rely on resolving interfaces, robust to complex surface morphology.

The test cases we present in Section \ref{sec:dropletResults} demonstrate that this method should be applicable to a broad range of droplet-related problems on substrates of interest, at least for sessile droplets and low-inertial flows.
In future work, we intend to model the chemical heterogeneity of plant leaf surfaces by varying the adhesive parameter $s_\mathrm{fs}$ across the surface to more accurately reflect the real surface chemistry.
Other potential applications include the study of spreading and impaction on rough or patterned surfaces.
With \firstRevision{the careful parameter calibration developed here}, the pairwise force model shows promise for simulating droplets on otherwise challenging substrates.